\documentclass[aps,physrev,twocolumn,preprint,floatfix,longbibliography,10pt,nofootinbib]{revtex4-2}
\makeatletter
\def\l@subsubsection#1#2{}
\makeatother
\usepackage[english]{babel}
\usepackage[utf8]{inputenc}
\usepackage[T1]{fontenc}
\usepackage{microtype}
\usepackage{amsmath,amssymb,amsthm,braket,slashed,mathrsfs}

\usepackage{pifont}

\usepackage[squaren,Gray,cdot,binary]{SIunits}
\usepackage{graphicx}
\graphicspath{{./figures/}}
\usepackage{color,hyperref}
\usepackage{cleveref}
\usepackage{marginnote}

\renewcommand{\vec}{\mathbf}

\crefformat{section}{Sec.~#2#1#3}
\crefmultiformat{section}{Secs.~#2#1#3}
{ and~#2#1#3}{, #2#1#3}{ and~#2#1#3}
\crefrangeformat{section}{Secs.~#3#1#4\,--\,#5#2#6}
\crefformat{subsection}{Sec.~#2#1#3}
\crefmultiformat{subsection}{Secs.~#2#1#3}
{ and~#2#1#3}{, #2#1#3}{ and~#2#1#3}
\crefrangeformat{subsection}{Secs.~#3#1#4\,--\,#5#2#6}
\crefformat{subsubsection}{Sec.~#2#1#3}
\crefmultiformat{subsubsection}{Secs.~#2#1#3}
{ and~#2#1#3}{, #2#1#3}{ and~#2#1#3}
\crefrangeformat{subsubsection}{Secs.~#3#1#4\,--\,#5#2#6}
\crefformat{figure}{Fig.~#2#1#3}
\crefmultiformat{figure}{Figs.~#2#1#3}
{ and~#2#1#3}{, #2#1#3}{ and~#2#1#3}
\crefrangeformat{figure}{Figs.~#3#1#4\,--\,#5#2#6}
\crefformat{table}{Tab.~#2#1#3}
\crefmultiformat{table}{Tabs.~#2#1#3}
{ and~#2#1#3}{, #2#1#3}{ and~#2#1#3}
\crefrangeformat{table}{Tabs.~#3#1#4\,--\,#5#2#6}
\crefformat{equation}{Eq.~(#2#1#3)}
\crefmultiformat{equation}{Eqs.~(#2#1#3)}{ and~(#2#1#3)}{, (#2#1#3)}{ and~(#2#1#3)}
\crefrangeformat{equation}{Eqs.~(#3#1#4)--(#5#2#6)}
\crefformat{appendix}{App.~#2#1#3}
\crefmultiformat{appendix}{Apps.~#2#1#3}
{ and~#2#1#3}{, #2#1#3}{ and~#2#1#3}
\crefrangeformat{appendix}{Apps.~#3#1#4\,--\,#5#2#6}

\usepackage{xcolor}
\definecolor{knorange}{HTML}{F39019}

\definecolor{linkcolor}{rgb}{.17578125,.1875,.5703125}
\hypersetup{
  pdfstartview={FitH},
  pdftitle={Decorrelating lattice correlators},
  pdfauthor={Antonin Portelli and Justus Tobias Tsang},
  pdfsubject={Physics research article},
  pdfcreator={latex},
  pdfkeywords=,
  colorlinks=true,
  linkcolor=linkcolor,
  citecolor=linkcolor,
  filecolor=black,
  urlcolor=linkcolor
}

\newcommand{\cf}{cf.}
\newcommand{\eg}{e.g.}
\newcommand{\ie}{i.e.}

\newcommand{\diff}{\mathrm{d}}

\renewcommand{\epsilon}{\varepsilon}
\newcommand{\F}{\mathcal{F}}

\DeclareMathOperator{\Prob}{P}

\newcommand{\dB}{\deci\bel}
\newcommand{\matrixel}[3]{\left< #1 \vphantom{#2#3} \right| #2 \left| #3 \vphantom{#1#2} \right>}

\begin{document}
\title{Application of Laplace filters to the analysis of lattice time correlators}
\author{Antonin Portelli}
\affiliation{School of
Physics and Astronomy, The University of Edinburgh, Edinburgh EH9 3FD, UK}
\author{Justus Tobias Tsang}
\affiliation{CERN, Department of Theoretical Physics}
\affiliation{Theoretical Physics Division, Department of Mathematical Sciences, University of Liverpool, Liverpool L69 3BX, UK}
\begin{abstract}
  The analysis of lattice simulation correlation function data is notoriously hindered by the ill-conditioning of the Euclidean time covariance matrix. Additionally, the isolation of a single physical state in such functions is generally affected by systematic contamination from unwanted states. In this paper, we present a new methodology based on regulated Laplace filters and demonstrate that it can be used to address both issues using state-of-the-art simulation data. Regulated Laplace filters are invertible high-pass filters that suppress local correlations in the data, and we show that they can reduce the condition number of covariance matrices by several orders of magnitude. Furthermore, Laplace filters can annihilate functions that decay exponentially with time, which can be used to alter the spectrum of a lattice correlation function. We show that this property can be exploited to significantly reduce excited-state contamination in the determination of matrix elements. The same property can also be used to constrain the spectral content of a correlation function and has the potential to form the basis of new methods to extract physical information from lattice data.
\end{abstract}
\maketitle

\section{Introduction}
Lattice quantum chromodynamics (QCD), or more generally, lattice field theory, allows one
to formulate strongly coupled field theories nonperturbatively. Such a formulation is then
often used as the basis for numerical ab initio simulations, which provide systematically
improvable predictions in particle physics and related fields. Compelling examples of
results obtained using lattice QCD can be found in the Flavour Lattice Averaging Group
(FLAG) review~\citep{FlavourLatticeAveragingGroupFLAG:2024oxs} and the Muon $g-2$ Theory
Initiative White Paper~\citep{Aliberti:2025beg}. In lattice QCD, a discretized version of
the QCD path integral in Euclidean space-time is numerically estimated using Markov chain
Monte Carlo (MCMC) techniques. The quantities typically computed are \emph{correlation
functions}, \ie, expectation values of products of local quantum operators. They
are generally expressed as functions of the momenta and Euclidean times of the
operators involved (time-momentum representation). Assuming all time separations between
operators are fixed except one, and assuming a theory with a mass gap $M>0$, correlation functions are  linear combinations of
exponentials of the form $e^{-Et}$, where $t$  is the variable Euclidean time separation,
and $E\geq M$ is the energy of a specific quantum state propagating in the system. The
coefficients (or \emph{amplitudes}) of such combinations are related to quantum matrix
elements of the operators involved. The energies and matrix elements are physical
quantities typically required for predicting a given observable.

A critical step in most lattice analyses is therefore to fit statistical data for correlation functions to 
a
truncated spectral decomposition as described above. This is generally done through a goodness-of-fit
optimization of such a model to the data. Although this is in appearance a simple problem, precisely assessing systematic uncertainties on the resulting fit parameters can be challenging, in
part due to the two issues below:
\begin{enumerate}
  \item Data points are generally strongly statistically correlated at short time
    separations, as a consequence of the covariance itself being a correlation function
    decaying exponentially with time separations. This makes $\chi^2$ goodness-of-fit
    estimates potentially unreliable due to poor conditioning of the time-covariance
    matrix.
  \item Numerous physical observables require the extraction of a single term $A\,e^{-Et}$
    in the correlation function, typically the one with the smallest energy (ground-state
    contribution). It is often difficult to distinguish such a term from others
    in the correlation function, and maximum-likelihood estimators of $A$ and $E$ can be
    systematically biased by the presence of contributions with energies close to $E$.
    This issue is generally referred to in the literature as \emph{excited-state contamination}.
    Additionally, in many lattice correlation functions, the signal-to-noise ratio
    degrades exponentially fast at large times, which can make this issue extremely
    challenging to address in a controlled way.
\end{enumerate}
These two issues can combine and result in uncontrollable systematic errors. Indeed,
excited-state contamination could, in theory, be addressed by conducting likelihood-ratio
tests for the presence of additional states; however, this requires reliable $p$-values
from minimized $\chi^2$, which is generally hindered by the poor conditioning of the
covariance matrix.

The issues above, as well as potential mitigation strategies, have been discussed at
length in the lattice QCD literature over the past 30 years. Some original work on the topic can be found in Refs.~\cite{Michael:1993yj,Michael:1994sz}, and here we review some of the most recent works on this topic. One approach to addressing the first issue above is
to use the better-behaved \emph{Ledoit-Wolf shrinkage}
estimators~\citep{ledoitWellconditionedEstimatorLargedimensional2004} for the covariance
matrix. Such an estimator replaces the sample covariance matrix $V$ with a regulated
version of the form $V^*=(1-\lambda)V+\lambda V_{\mathrm{diag}}$, where
$V_{\mathrm{diag}}$ is the diagonal part of $V$. It can be
proven~\citep{ledoitWellconditionedEstimatorLargedimensional2004} that the regulator
$\lambda$ can be chosen such that, asymptotically, $V^*$ is a better-conditioned unbiased
estimator of the covariance matrix. This approach is commonly used in lattice QCD; see,
for example,~\citep{Green:2017keo} as one of the earliest works to use it. A key issue
with this approach is that although $V^*$ can be shown to be an optimal regularization
asymptotically, it may be difficult to estimate the deviation from this behavior at finite
statistics. The shrinkage described above was then improved using a non-linear optimization in Ref.~\cite{10.1214/12-AOS989,Ledoit_Wolf_2018}. However, without studying the asymptotic properties, Ledoit-Wolf shrinkage becomes
essentially an ad hoc regularization procedure, and it is difficult to assess its effect
on estimating $p$-values and resulting hypothesis tests.

Another approach proposed in~\citep{Bruno:2022mfy} suggests using an alternative statistic
to the standard correlated $\chi^2$, where the covariance matrix is replaced by a
better-behaved one, typically the diagonal part of the covariance matrix. This formalism covers other common
regularization strategies such as cuts in the singular value decomposition of the
correlation matrix (\cf~for example Ref.~\citep{Dowdall:2019bea}). In this work, the authors
derive the appropriate $p$-value for this modified $\chi^2$, allowing them to perform
hypothesis tests based on a more reliable estimator. While this addresses some
shortcomings of more ad hoc regularization strategies, the fit parameters obtained by
minimizing the modified $\chi^2$ are, strictly speaking, not the maximum-likelihood
parameters given the correlated distribution of the data. 

Finally, regarding the determination of a single-state contribution (issue 2 above), a
number of techniques are commonly employed. First, excited-state contamination can be
suppressed at the level of the measurement itself by choosing operators with weaker
coupling to the unwanted states. This can be achieved, for example, by smearing operators
using Jacobi iterations~\citep{UKQCD:1993gym} or more general forms targeting specific
physics (see, for example, the recent work~\citep{Knechtli:2022bji}). Methods based on
operator design may not be confined to purely statistical data analysis and may require
re-measuring expensive correlation functions.

Restricting ourselves to methods usable at
the data-analysis stage, one can attempt to use the time-translation properties of
correlation functions to reconstruct the spectrum. This is historically the aim of Prony's
method~\citep{Prony}, which was adapted to lattice QCD analysis as a generalized
eigenvalue problem (GEVP) in~\citep{Fleming:2004hs,Beane:2009kya}. More recently, Wagman
proposed a new method~\citep{Wagman:2024rid} using time translations of the correlation
function to infer the spectrum of the transfer matrix via the Lanczos algorithm. This was proven
later~\citep{Ostmeyer:2024qgu} to be equivalent to the Prony GEVP, and this equivalence is
further discussed in~\citep{Abbott:2025yhm}. Although methods involving solving for a
matrix spectrum have been used consistently in lattice QCD~\citep{Blossier:2009kd}, they
can feature challenging eigenvalue-identification issues due to the noisy nature of the
data. There are ways to regularize this problem, including in the previously cited
references, or additionally in the recent work~\citep{Chakraborty:2024exj}.

In this paper, we propose a new method to solve the two issues above, with an emphasis on
simplicity and minimal deviations from standard statistical frameworks. We stress that this method is complementary to the literature reviewed in this section, in the sense that in cases where it does not by itself overcome the mentioned issues, it can be applied in conjunction with other ways to regularise the resulting covariance matrix. The key concept is
to use regulated Laplace filters on correlation functions, which are a discretized version
of the second-order differential operator $-\partial^2_t+\lambda^2$, where $\lambda$ is an
arbitrary real, nonzero regulator that guarantees the invertibility of the filter. Laplace
filters are high-pass filters, allowing us to decorrelate data and improve the
conditioning of the correlation matrix. Additionally, they annihilate $e^{\pm\lambda t}$
at nonzero times, allowing us to modify and probe the spectrum of a correlation function
by varying the regulator. In \cref{sec:general}, we give general definitions that will be
used in the rest of the paper. In \cref{sec:data}, we introduce representative lattice QCD
data that will be used for numerical tests of our methods. In \cref{sec:laplace}, we
discuss how Laplace filters can be used to significantly improve the conditioning of
covariance matrices. We then demonstrate in \cref{sec:spectrum} how statistical constraints on
the spectrum of a correlation function can be designed using Laplace filters. Finally, we
present in \cref{sec:application} a number of applications to realistic lattice QCD
data-analysis examples.

\section{General problem and definitions}
\label{sec:general}
We consider space-time to be a finite discrete hyper-cubic lattice.\footnote{This is the most straightforward and common choice in lattice field theory, however every method in this paper can be applied to anisotropic lattices.} We consider an arbitrary quantum expectation value lattice with spacing $a$ and Euclidean
metric
\begin{equation}
  C[t]=\braket{O[t]}
\end{equation}
depending on a single discrete time separation $t$, which in units of the time spacing $a$
takes integer values from $0$ to $N_t-1$. The operator $O$ is generally a product of two or more local operators, and
may have additional time variables, which are assumed fixed here. Such an object is
referred to as a \emph{time correlator}. Unless otherwise specified, all quantities are
expressed in units of $a$. The bracket notation $[t]$ indicates that $t$ is a discrete
variable; later, the notation $(t)$ will be used for continuous variables. We also denote
by $C$ the column vector whose component at row $t$ is $C[t]$. We always assume that $N_t$
is divisible by $2$. Unless otherwise specified, we assume discrete times are periodic in
this paper. This means any operation on time indices should be understood as being
performed modulo $N_t$. This assumption is, in principle, arbitrary, and in several cases
lattice correlators may be computed with non-periodic boundary conditions in time. We make
this choice primarily for clarity, as several concepts in this paper have known
generalizations to non-periodic boundary conditions. Secondly, a finite sequence of
numbers can always be assumed periodic regardless of its origin, and the boundary
condition can be seen as potential additional freedom in several methods presented here.

Generally, $C$ is estimated through a Monte Carlo process, and $N$ samples $C_{\alpha}[t]$
of $C[t]$ are known, either directly or through a resampling process (e.g., jackknife or
bootstrap). It is important to note that we assume $\alpha$ indexes samples of the
expectation value $C$, not the value of the underlying operator $O$ on individual field
configurations\footnote{Assuming normally distributed averages, this assumption simply
means we do not add a factor of $\frac{1}{N}$ in the sample variance matrix definition in \cref{eq:covmat}.}.
The average
\begin{equation}
  \bar{C}[t]=\frac{1}{N}\sum_{\alpha=1}^N C_{\alpha}[t]\,,
\end{equation}
is a Monte Carlo estimator of $C[t]$. We define the associated noise function:
\begin{equation}
  \eta_\alpha[t]=C_\alpha[t]-\bar{C}[t]\,.
\end{equation}
By construction, $\bar{\eta}[t]=0$. An unbiased estimator of the true time
\emph{covariance matrix} $\Sigma[t_1,t_2]$ of $C[t]$ is given by the \emph{sample
covariance matrix}
\begin{align}
  V[t_1,t_2]&=\frac{N}{N-1}\bigl(\overline{\eta\otimes\eta}\,[t_1,t_2]\bigr)\notag\\
  &=\frac{1}{N-1}\sum_{\alpha=1}^N\eta_{\alpha}[t_1]\eta_{\alpha}[t_2]\,.
  \label{eq:covmat}
\end{align}
and the \emph{sample correlation matrix} is defined as
\begin{equation}
  \Gamma[t_1,t_2]=\frac{V[t_1,t_2]}{\sqrt{V[t_1,t_1]V[t_2,t_2]}}\,.
\end{equation}
Both $V$ and $\Gamma$ are symmetric, positive-definite matrices. The correlation matrix
can also be defined analogously to~\cref{eq:covmat} using the \emph{normalized noise
function}:
\begin{equation}
  \gamma_{\alpha}[t]=\frac{\eta_\alpha[t]}{\sqrt{V[t,t]}}\,,
\end{equation}
so that $\Gamma$ can be written:
\begin{equation}
  \Gamma[t_1,t_2]=\frac{N}{N-1}\bigl(\overline{\gamma\otimes\gamma}\,[t_1,t_2]\bigr)\,.
\end{equation}

In a typical lattice calculation, one fits $\bar{C}[t]$ to a model $f(t;\theta)$,
where $\theta$ denotes the parameters of the model. This model is generally determined by a
spectral representation of the underlying correlation function. This is often achieved using a $\chi^2$ goodness-of-fit approach as follows. We define the model prediction vector
$F(\theta)[t]$ as the column vector with coefficients $f(t;\theta)$, and the model-to-data
variation $\Delta(\theta)=F(\theta)-\bar{C}$. The $\chi^2$ function for the fit is then
given by
\begin{equation}
  \chi^2(\theta)=\Delta(\theta)^{T}V^{-1}\Delta(\theta)\,.
  \label{eq:chi2}
\end{equation}
Maximum-likelihood parameters are then found by minimizing $\chi^2(\theta)$ over $\theta$. A key
issue in this process is the presence of the inverse covariance matrix in~\cref{eq:chi2}.
In a system with strong time correlations, which is often the case with lattice QCD data,
$V$ can be badly conditioned, resulting in a numerical inverse that is poorly determined
or even impossible to compute with double-precision arithmetic. In such cases, maximum-likelihood
parameters or the $\chi^2$ function itself can have unreliable values, compromising the
scientific output of the calculation. Additionally, it has become frequent in
state-of-the-art lattice calculations to estimate final systematic uncertainties using a
model-averaging procedure~\citep{BMW:2014pzb,Jay:2020jkz}. In such an approach, many
candidate predictions for a given observable are combined in a histogram weighted
according to some score function, for example the Akaike information criterion (AIC).
Score functions will often depend on numerical values of $\chi^2$, and imprecision in
estimating those can have non-trivial effects on the final prediction.

The problem of inverting the covariance matrix can be replaced without loss of generality
by the invertibility of the correlation matrix, which is in general better conditioned. We
define $D[t]=V[t,t]^{-\frac12}$ to be the element-wise inverse square root of the diagonal
vector of the covariance matrix. The correlation matrix $\Gamma$ is then given by
\begin{equation}
  \Gamma=\mathrm{diag}(D) \,V\, \mathrm{diag}(D)\,,
\end{equation}
where $\mathrm{diag}(D)$ is the diagonal matrix with coefficients given by the
elements of $D$. Therefore, knowing $(D\otimes D)$ from the data, inverting $V$ can be
reduced to inverting $\Gamma$, whose coefficients are bounded in the $[-1,1]$ interval. We
now discuss a metric that can be used to quantify the conditioning of $\Gamma$.
\subsection{Correlation dynamic range (CDR)}
\label{sec:cdr}
We introduce the notion of
\emph{correlation dynamic range} (CDR). We first define the \emph{condition number} of a
positive-definite matrix $M$, given in the usual way by
\begin{equation}
  \kappa(M)=\frac{\max_j(\sigma_j)}{\min_j(\sigma_j)}\,,\label{eq:condition}
\end{equation}
where the $\sigma_j$, $0\leq j < N_t$, are the eigenvalues of $M$. The CDR of $M$ is then
defined by the value of $\kappa(M)$ in decibels ($\dB$), namely
\begin{equation}
  \mathrm{CDR}(M)=10\log_{10}[\kappa(M)]\,.\label{eq:cdr}
\end{equation}
The CDR gives the number of orders of magnitude between the smallest and largest
eigenvalues of $M$, multiplied by $10$. A general rule of thumb in interpreting the CDR is
as follows: a CDR of $x~\dB$ implies that the precision of the data will be degraded by
$x/10$ significant digits when multiplied by $M$.

In the case of the correlation matrix $\Gamma$, exactly uncorrelated data is equivalent to
$\Gamma=1$, which is equivalent to a CDR of $\unit{0}{\dB}$. On the other hand, if one
data point is an exact linear combination of others, then $\Gamma$ is singular and has a
CDR of $\unit{+\infty}{\dB}$. With IEEE 754 double precision arithmetic, a matrix becomes
effectively singular for CDRs above approximately
$10\log_{10}(2^{53})\simeq\unit{156.54}{\dB}$. Let us now define the general problem of
optimization of correlations through data transformations.
\subsection{General data transformations}
\label{sec:data-trans}
\noindent Let $\phi$ be an arbitrary known function from the data space $\mathbb{R}^{N_t}$
to some \emph{latent space} $\mathbb{R}^{N_\ell}$, with $N_\ell\leq N_t$. We define the
latent-space data
\begin{equation}
  C_{\alpha}^{(\phi)}=\phi(C_{\alpha})\in\mathbb{R}^{N_\ell}\,,
\end{equation}
and we denote with a superscript $(\phi)$ all quantities derived from $C_\alpha$ (\eg~
$\bar{C}$, $\eta_{\alpha}$, $V$, etc.) as the corresponding estimators evaluated on the
latent-space data $\smash{C_{\alpha}^{(\phi)}}$. As defined in data space
for~\cref{eq:chi2}, given a model $f(t;\theta)$, we define the latent-space model
prediction vector
\begin{equation}
  F^{(\phi)}(\theta)=\phi[F(\theta)]\,,
\end{equation}
where, as previously, $F(\theta)[t]=f(t;\theta)$. One can hence search for best-fit
parameters by minimizing the latent-space $\chi^2$ function
\begin{equation}
  (\chi^2)^{(\phi)}(\theta)=\Delta^{(\phi)}(\theta)^{T}(V^{(\phi)})^{-1}\Delta^{(\phi)}(\theta)\,,
\end{equation}
with $\Delta^{(\phi)}(\theta)=F^{(\phi)}(\theta)-\bar{C}^{(\phi)}$. One can then attempt
to find transformations under which, for example, the correlation matrix is better
conditioned. We also require that the model
predictions and the $\chi^2$ function remain meaningful from a statistical point of view.
Therefore, the latent-space data should approximately obey Gaussian statistics for the
$\chi^2$ function interpretation to be valid. In general, an important special case is
when $\phi$ is a linear transformation, which will always be the case in this paper. Under
this assumption, if the data are Gaussian distributed, then the latent-space data will
also be Gaussian with mean and variance given by
\begin{equation}
  \bar{C}^{(\phi)}=\Phi\bar{C},\qquad\text{and}\qquad
  V^{(\phi)}=\Phi V\Phi^T\,,
\end{equation}
where $\Phi$ is the $N_\ell\times N_t$ matrix representation of $\phi$. If $\phi$ is
nonlinear, then the Gaussianity of the latent-space data will directly depend on the
smoothness of $\phi$ around $\bar{C}$.

In this paper, we will investigate a unique class of linear transformation, the regulated
Laplace Filters, presented in the next section. However, the general idea in this section
may be of use for future research on different transformations.
\subsection{Regulated Laplace Filter definition}
\label{sec:filter-def}
\noindent
We start by considering the discrete Laplacian
\begin{equation}
  \Delta C[t]=C[t-1]-2\,C[t]+C[t+1]\,.
\end{equation}
The data transformation considered here is the operator
\begin{equation}
  D_\lambda C[t]=(-\Delta+\lambda^2)\,C[t]\,,
  \label{eq:laplace}
\end{equation}
where the regulator $\lambda$  is a nonzero real number and ensures that $D_\lambda$ is an
invertible matrix. $D_\lambda$ is translation-invariant and corresponds to the convolution
with the impulse response $(-1,2+\lambda^2,-1)$. This filter will perform local
subtractions in time, which intuitively should reduce correlations in lattice time
correlators. In frequency space, this corresponds to the multiplication
\begin{equation}
  \F D_\lambda C[\omega]=\left[4\sin\left(\frac{\pi\omega}{N_t}\right)^2
  +\lambda^2\right]\hat{C}[\omega]\,,
\end{equation}
where $\F C=\hat{C}$ is the discrete Fourier transform of $C$:
\begin{equation}
  \hat{C}[\omega]=\sum_{t=0}^{N_t-1}C[t]\,e^{-i\frac{2\pi}{N_t} \omega t},
  \qquad\text{with}\quad \omega\in\{0,\dots,N_t-1\}\,.
\end{equation}
In essence, $D_\lambda$ is a square-law high-pass filter, which suggests it will suppress
short-time correlations. Before discussing the practical implication of this
statement further, we present the representative lattice QCD data used in this paper.
\section{Representative sample data}
\label{sec:data}
\begin{table}
  \begin{ruledtabular}
    \begin{tabular}{lcccc}
      name & $M_\pi/\mathrm{MeV}$ & $a^{-1}/\mathrm{GeV}$ & flavor & process\\\hline
      M0M-mes-ll  & 139 & 2.3 & $\bar{\ell}\ell$ & $\pi\to \pi$\\
      F1M-mes-lh  & 232 & 2.7 & $\bar{\ell}h$ & $D\to D$\\
      F1M-mes-hh  & 232 & 2.7 & $\bar{h}h$ & $\eta_c\to \eta_c$\\
      C1M-semi-sc-ss & 276 & 1.7 & $\bar{s}c\to\bar{s}s$& $D_s \to \eta_s \ell \nu$\\
      F1M-mix-sh  & 232 & 2.7 & $s\bar{h}\to\bar{s}h$ & $B^0_s\to \overline{B}_s^0$\\
    \end{tabular}
  \end{ruledtabular}
  \caption{Representative datasets used in this study. The temporal extents
    $N_t$ of the M0M, F1M and C1M ensembles are 128, 96 and 64,
  respectively.\label{tab:data}}
\end{table}
In order to exemplify methods in this paper, we apply them to real lattice data on $N_f=2+1$
ensembles generated by the RBC and UKQCD collaborations~\citep{RBC:2014ntl,
Boyle:2018knm, Boyle:2024gge}. We consider five datasets corresponding to a
selection of typical state-of-the-art lattice applications. In
\cref{tab:data}, we list some basic properties of these datasets. In particular, we
consider mesonic pseudoscalar two-point functions: one for a pion with physical pion
mass (M0M-mes-ll), one for a light-heavy meson between the $D$ and the $B$ masses
(F1M-mes-lh), and one for a heavy-heavy meson between the $\eta_c$ and the $\eta_b$
masses (F1M-mes-hh). Later, we also consider two-point and three-point
functions related to the semileptonic decay of a charm quark to a strange quark
with a strange spectator quark (C1M-semi-sc-ss), and to neutral meson mixing
between a neutral, unphysically light $B_s$ meson and its antiparticle
(F1M-mix-sh). While this is not an exhaustive list, it nonetheless
captures several types of datasets to which the ideas presented here can be applied
and which are directly relevant to the phenomenological exploitation of lattice
QCD simulations.

Regarding the M0M-mes-ll, F1M-mes-lh, and F1M-mes-hh correlation functions, we consider
standard operators with the quantum numbers of pseudoscalar mesons of the form
\begin{equation}
  \phi_{q_1q_2}(x) = \left[ \bar{q}_2 \gamma_5 q_1\right] (x)\,,
\end{equation}
where $q_1$ and $q_2$ are quark fields of potentially different flavors.
The correlators studied are two-point functions of the form
\begin{equation}
  C^{q_1q_2}_2[t] = \sum_{\vec{x}, \vec{y}} \braket{\phi_{q_1q_2}(y)
  \phi_{q_1q_2}^\dagger(x)}\,,
\end{equation}
with $x=(t_\mathrm{src},\vec{x})$, $y=(t_\mathrm{snk},\vec{y})$, and 
$t = t_\mathrm{snk} - t_\mathrm{src}$.
Also, we only consider zero-momentum-projected correlation
functions, so we do not list any momentum arguments, even though
results in this paper equally apply to nonzero momenta.

For the semileptonic decay dataset C1M-semi-sc-ss, we additionally consider the temporal
component of the local vector current
\begin{equation}
  V_{q_1q_2}(x) = \left[\bar{q}_1 \gamma_0 q_2\right](x)
\end{equation}
and, identifying $t_\mathrm{snk} = z_0$, $t_\mathrm{op} = y_0$, and $t_\mathrm{src}
= x_0 = 0$, the correlation functions
\begin{equation}
  C^{q_1q_3\to q_2q_3}_3[t_\mathrm{op},t_\mathrm{snk}] =
  \sum_{\vec{x},\vec{y},\vec{z}}\braket{\phi_{q_2q_3}(z) V_{q_1q_2}(y)
  \phi_{q_1q_3}^\dagger(x)}\,.
\end{equation}
Finally, for the mixing data set F1M-mix-sh, we consider the Standard Model four-quark
bag parameter operator commonly referred to as $O_1 = O_{VV+AA}$ defined as
\begin{equation}
  O_1^{q_1q_2} =(\bar{q}_2 \gamma_\mu q_1) \, (\bar{q}_2 \gamma_\mu q_1)
  + (\bar{q}_2 \gamma_\mu\gamma_5 q_1) \, (\bar{q}_2 \gamma_\mu\gamma_5 q_1)\,,
\end{equation}
where summation over the index $\mu$ is implicit.
The corresponding three point functions have the form
\begin{equation}
  C^{q_1q_2}_3[t_\mathrm{op},t_\mathrm{snk}] = \sum_{\vec{x},\vec{y},\vec{z}}
  \braket{ \phi_{q_2q_1}(z) O_1^{q_1q_2}(y) \phi_{q_1q_2}^\dagger(x)}\,,
\end{equation}
where we again identify $t_\mathrm{snk} =z_0$, $t_\mathrm{op}=y_0$ and
$t_\mathrm{src} = x_0 = 0$.  In all cases the spatial sums are performed
stochastically using $\mathbb{Z}_2$-wall sources and local sinks. In the following we
give some technical detail about the generation of the data sets, in each case
first giving the flavor assignment.
\begin{description}
  \item[M0M-mes-ll] The flavor assignment is $q_1 = \ell$, $q_2 = \ell$ where
    $\ell$ is a physical light quark. For this particular correlation function,
    the sources were additionally smeared using Jacobi
    iterations~\citep{UKQCD:1993gym}. For further
    technical details about this data set we refer the reader to
    Refs.~\citep{Boyle:2018knm, Boyle:2024gge}.
  \item[F1M-mes-lh] Here we assign $q_1 = \ell$, $q_2 = h$ where the light quark
    $\ell$ takes the unitary value for this ensemble and the heavy quark mass $am_q =
    0.68$ lies between the physical charm and bottom quark mass. For the heavy
    quark we use one of the actions compared in Ref.~\citep{Cho:2015ffa}. More
    details about the exact parameter choices are given in
    Ref.~\citep{Boyle:2018knm} where this dataset was first introduced.
  \item[F1M-mes-hh] This is the same as F1M-mes-lh, but with $q_1 = h$ and $am_h =
    0.59$ for both quarks. Only the quark-connected part of this correlation
    function has been considered.
  \item[C1M-semi-sc-ss] The assignment of flavors for this dataset is $q_1 =
    c$, $q_2 = s$, $q_3 = s$. The computational set-up is similar
    to that of Ref.~\citep{Flynn:2023nhi}, but using $Z_2$-wall sources instead of
    point sources: At the source propagators for a spectator strange quark ($q_3$)
    and a child strange-quark ($q_1$) are generated, at time $t_\mathrm{snk}$ the
    spectator quark is turned into a sequential source from which the parent
    charm-quark ($q_2$) is inverted. The charm quark mass is tuned to its physical
    value and the actions are the same as in the previous dataset. Measurements took
    place on 180 configurations with 8 sources per configuration.
  \item[F1M-mix-sh] We are considering $q_1 = s$, $q_2 = h$ here with $am_h =
    0.59$. This dataset has also been introduced in Ref.~\citep{Boyle:2018knm} and
    the computational set-up is the same as that for F1M-mes-lh.
\end{description}

\section{Improvement of time correlations}
\label{sec:laplace}
\begin{figure*}[t]
  \centering
  \includegraphics{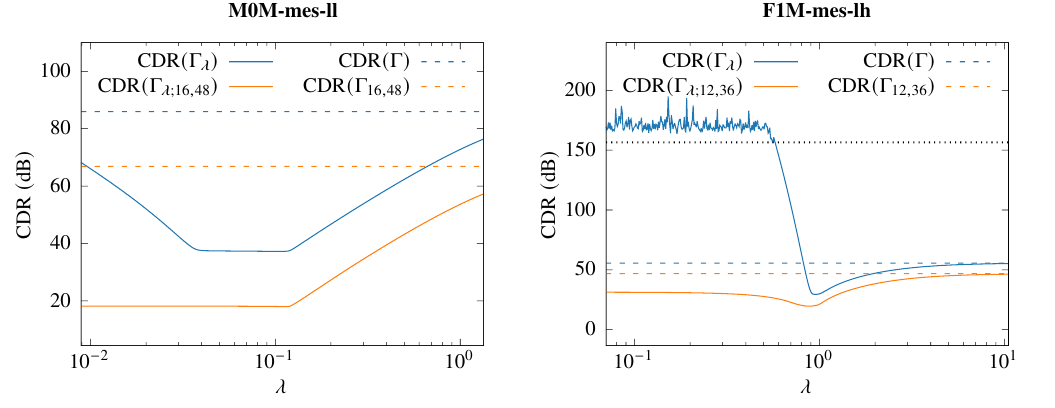}
  \caption{CDR of the filtered correlators as a function of the regulator $\lambda$. The
    left and right panels show the results for the M0M-mes-ll and F1M-mes-lh correlators,
    respectively. In each plot, the solid blue and orange curves represent the CDR for
    the full time range and for the restricted range $\{N_t/8,\ldots,3N_t/8-1\}$,
    respectively. Additionally, the colored dashed lines represent the CDRs of the
    unfiltered data, and the black dashed line represents the CDR of the restricted range
    evaluated at the optimal value of the regulator, with values of approximately
    $\lambda\simeq0.11$ for M0M-mes-ll and $\lambda\simeq0.88$ for F1M-mes-lh. In the
    F1M-mes-lh plot, the dotted line indicates the critical \unit{156.54}{\dB} threshold above
    which the CDR cannot be resolved using double-precision arithmetic
  (cf.~\cref{sec:cdr}).}
  \label{fig:optim}
\end{figure*}
\begin{figure*}[p]
  \centering
  \includegraphics{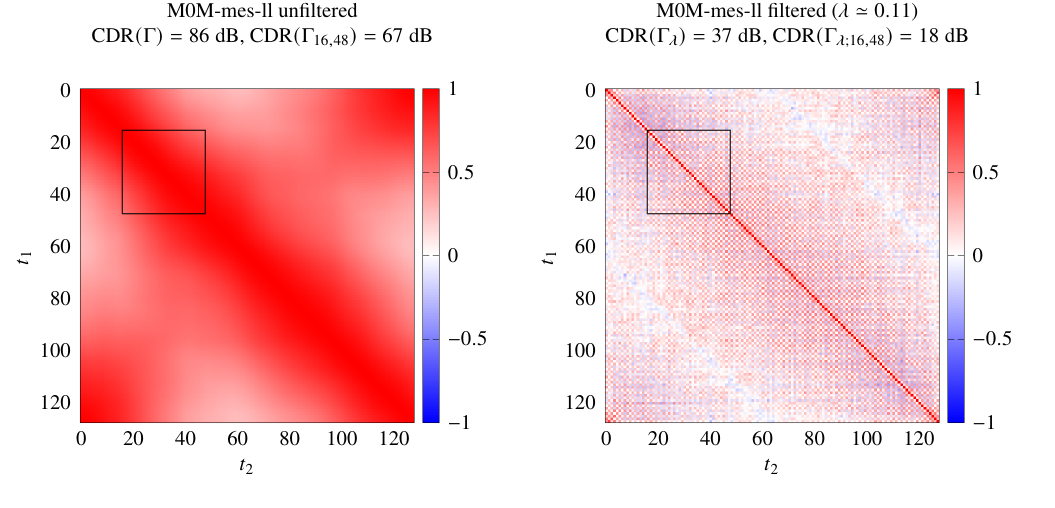}
  \caption{Correlation matrix of the M0M-mes-ll time correlator,
    in both the unfiltered (left panel) and filtered (right panel) cases. For the filtered
    case, the optimal value $\lambda\simeq0.11$ is used (cf.~\cref{fig:optim}).
    The titles of the plots provide the CDRs of the full and restricted
    matrices, using the same time range as in~\cref{fig:optim}. Finally, the restricted
    range is represented as a black square in the figure.}
  \label{fig:M0ll-corr}
\end{figure*}
\begin{figure*}[p]
  \centering
  \includegraphics{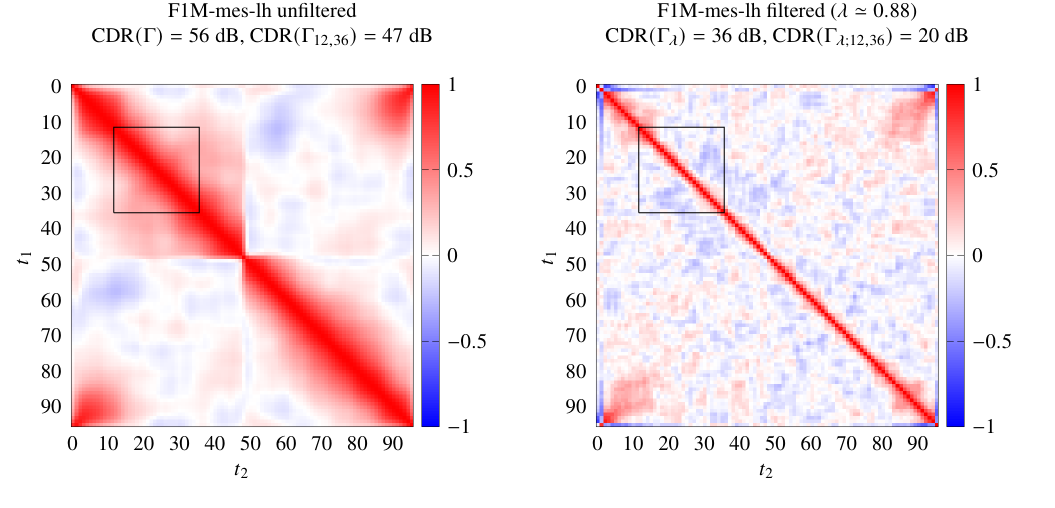}
  \caption{Representation of the correlation matrix of the F1M-mes-lh time correlator,
    in both the unfiltered (left panel) and filtered (right panel) cases. In the filtered
    case, the optimal value $\lambda\simeq0.88$ is used (cf.~\cref{fig:optim}). Other
    details are as in~\cref{fig:M0ll-corr}.}
  \label{fig:F1Mhl-corr}
\end{figure*}

The main question in this section is to
find values of $\lambda$ for which the CDR of the filtered data is
significantly reduced. This is in general a non-trivial question since the CDR depends on
the spectrum of the correlation matrix.

In practice, one rarely needs to fit the entire time range of a lattice correlator.
Instead, fits generally focus on a limited time range. It is
therefore useful to define a \emph{restriction operator} $R_S$, where $S$ is a subset of
$\{0,\dots,N_t-1\}$, that projects $C$ onto the components indexed by $S$. If $S$ is an
interval $\{t_i,\dots,t_f-1\}$, we define $R_{t_i,t_f}=R_S$. It is clear that $R_S$ is a linear
operator. We additionally define the following shorthand notation
\begin{align}
  \label{eq:restr1}
  &C_{\lambda}=D_{\lambda}C,\qquad C_{t_i,t_f}=R_{t_i,t_f}C,\\
  &\text{and}\qquad C_{\lambda;t_i,t_f}=R_{t_i,t_f}D_{\lambda}C\,\label{eq:restr2}.
\end{align}
It is important to note in \cref{eq:restr2} that the restriction operation is applied after
the Laplace filter.
We extend this notation to quantities that depend on $C$; for example,
$\Gamma_{\lambda}$ is the correlation matrix of $C_\lambda$, and so on.

The effect of $D_\lambda$ can be derived exactly in the limits of large and small
$\lambda$. For sufficiently large $\lambda$
the Laplace filter $D_\lambda$ is, to leading order, proportional
to the identity matrix and therefore has no effect on correlations; explicitly
\begin{equation}
  \lim_{\lambda\to+\infty}\mathrm{CDR}(\Gamma_{\lambda;t_i,t_f})
  =\mathrm{CDR}(\Gamma_{t_i,t_f})\,,
  \label{eq:liminf}
\end{equation}
for any time range from $t_i$ to $t_f$. On the other hand, when $\lambda\to 0^+$, $D_{\lambda}$
becomes non-invertible, resulting in a singular $\Gamma$:
\begin{equation}
  \lim_{\lambda\to 0^+}\mathrm{CDR}(\Gamma_{\lambda})=\unit{+\infty}{\dB}\,.
  \label{eq:limzero}
\end{equation}
However, the above limit is generally expected to remain finite if a nontrivial time
restriction is applied. We now discuss how to optimize the value of $\lambda$ 
to minimize correlations.
\subsection{Filter optimization}
\label{sec:filter-opt}
In this section, we perform an initial analysis of the effect of the regulated Laplace
filter on the statistical correlations in mesonic
two-point functions. More specifically, we consider the pseudoscalar correlators labeled M0M-mes-ll and F1M-mes-lh in~\cref{sec:data}. In both cases, we study how the CDR of the filtered
correlator varies as a function of $\lambda$. We compute the CDR for the full correlation
matrices and for the restricted range $\{N_t/8,\dots,3N_t/8-1\}$, which is viewed as a
representative fit range.

In all cases, the CDR is computed numerically using a Jacobi singular-value decomposition
in double precision. An optimal value of the regulator is then obtained by numerically
minimizing the filtered, restricted CDR as a function of $\lambda$. Once a solution
$\lambda_0$ is obtained, we sample 500 values of the CDR, uniformly spaced on a
logarithmic scale between $0.08\,\lambda_0$ and $12\,\lambda_0$. The results of this
analysis for both M0M-mes-ll and F1M-mes-lh are illustrated in~\cref{fig:optim}. As a
first observation, we see that the limits in~\cref{eq:liminf,eq:limzero} are satisfied. In
the case of the CDR divergence for $\lambda\to 0^+$, the CDR numerically saturates at the
maximum $\unit{156.54}{\dB}$ level for double-precision arithmetic. Secondly, in both the
M0M-mes-ll and F1M-mes-lh cases, there are values of $\lambda$ for which the filtered data
exhibit significantly less correlation than the unfiltered data. We also observe that the
optimal region is similar for both the full and restricted CDRs.
In~\cref{fig:M0ll-corr,fig:F1Mhl-corr}, we represent the correlation matrices of the
M0M-mes-ll and F1M-mes-lh correlators, as well as for the filtered data at the optimal
values $\lambda_0$. These figures also provide the numerical values of the full and
restricted CDRs for the original and filtered data. In the M0M-mes-ll case, the Laplace
filter improves both the full and restricted CDRs by $\unit{49}{\dB}$. In the F1M-mes-lh
case, the full CDR is improved by $\unit{20}{\dB}$, and the restricted CDR by
$\unit{27}{\dB}$. For all cases, a significant improvement is observed, and for the
restricted time range, the condition number of the filtered correlation matrix is at most
100, implying that it is well-conditioned.

Following these encouraging initial results, we discuss how additional restriction of the
data can further reduce the level of correlation.
\begin{table}[t]
  \begin{ruledtabular}
    \begin{tabular}{lllll}
      & \multicolumn{2}{c}{\textbf{M0M-mes-ll}}            &
      \multicolumn{2}{c}{\textbf{F1M-mes-lh}}             \\
      & total            & range            & total            & range            \\
      \hline
      $\mathrm{CDR}(\Gamma)$                          & $\unit{86}{\dB}$ &
      $\unit{67}{\dB}$ & $\unit{56}{\dB}$ & $\unit{47}{\dB}$ \\
      $\mathrm{CDR}(\Gamma_{\lambda_0})$              & $\unit{37}{\dB}$ &
      $\unit{18}{\dB}$ & $\unit{36}{\dB}$ & $\unit{20}{\dB}$ \\
      $\mathrm{CDR}(\Gamma^{\downarrow})$             & $\unit{62}{\dB}$ &
      $\unit{52}{\dB}$ & $\unit{37}{\dB}$ & $\unit{30}{\dB}$ \\
      $\mathrm{CDR}(\Gamma^{\downarrow}_{\lambda_0})$ & $\unit{21}{\dB}$ &
      $\unit{11}{\dB}$ & $\unit{19}{\dB}$ & $\unit{9}{\dB}$
    \end{tabular}
  \end{ruledtabular}
  \caption{Summary of CDR improvements obtained in~\cref{sec:laplace} for the M0M-mes-ll
    and F1M-mes-lh correlators. As assumed in all results of~\cref{sec:laplace},
    the range-restricted CDRs are evaluated over the representative time interval
    $\{N_t/8,\dots,3N_t/8-1\}$, and $\lambda_0$ denotes the regulator value that
  minimizes the range-restricted CDR. In principle, optimizing $\lambda_0$ for the
  total range can lead to lower CDR values in the ``total'' column.}
  \label{tab:cdr}
\end{table}

\subsection{Additional downsampling step}
\label{sec:ds}
\begin{figure*}[p]
  \centering
  \includegraphics{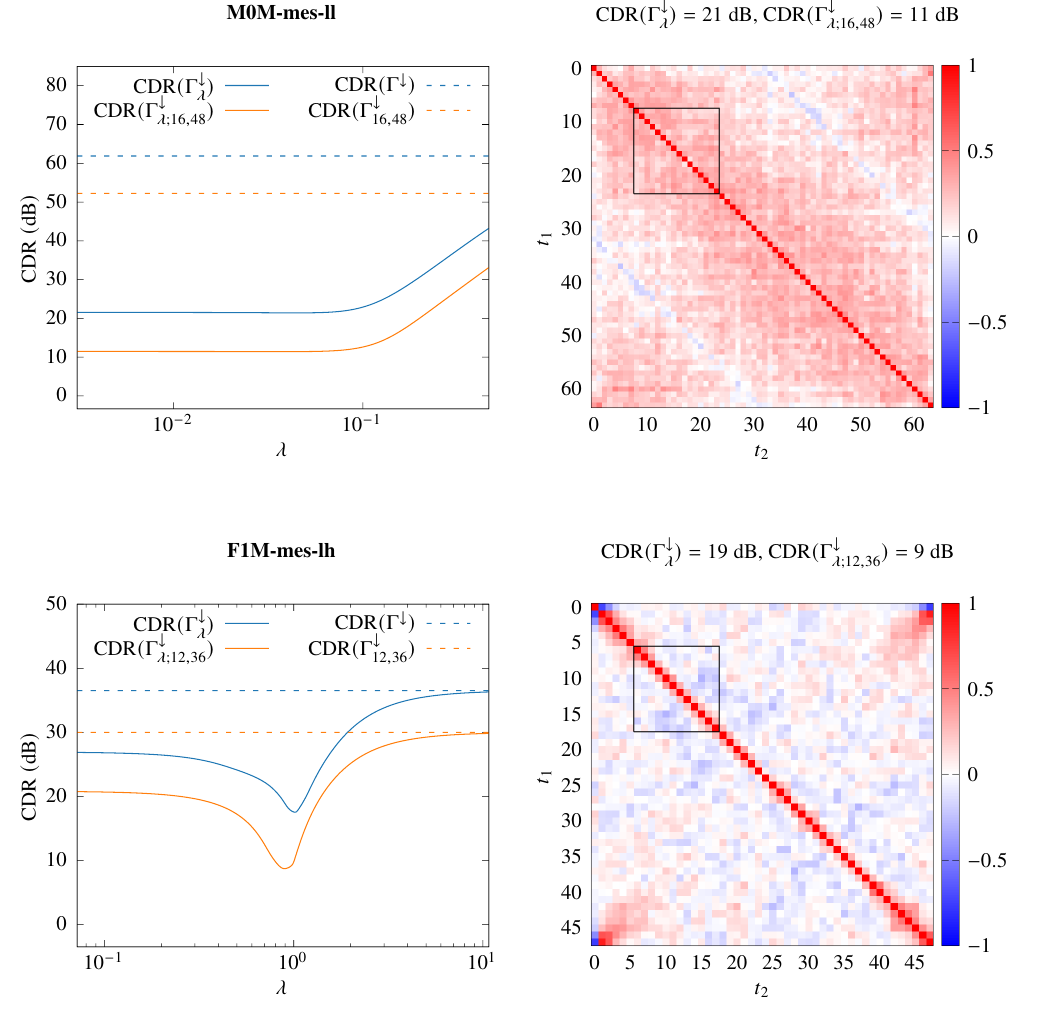}
  \caption{Laplace filter analysis and optimization when using an additional downsampling
    step after filtering, as explained in~\cref{sec:ds}. The two rows of plots are for the
    analysis of the M0M-mes-ll and F1M-mes-lh correlators, respectively. Plots on the left
    represent the dependence of the total and restricted CDR on the filter regulator
    $\lambda$, using identical conventions to~\cref{fig:optim}. Plots on the right
    represent the correlation matrix $\Gamma^{\downarrow}_{\lambda}$, for the optimal
    value $\lambda=\lambda_0$, which minimizes the restricted downsampled CDR (i.e., the
    orange curve on the associated plot on the left). The variables $t_1$ and $t_2$ are time indices of the downsampled correlator; that is, they correspond to $2t_1$ and $2t_2$ for the original data, respectively. As
    in~\cref{fig:M0ll-corr,fig:F1Mhl-corr}, the sub-matrix within the black square is the
  correlation matrix on the restricted range.}
  \label{fig:ds}
\end{figure*}
We define the \emph{downsampled} or \emph{thinned} correlator $C^{\downarrow}$,
of size $N_t/2$, as follows:
\begin{equation}
  C^{\downarrow}[t]=C[2t]\,,
\end{equation}
\ie, all odd time slices of the original correlator are discarded. Equivalently,
$C^{\downarrow}=R_EC$, where $E$ is the set of even time slices. The choice of discarding
specifically odd time slices is arbitrary and made here for the sake of simplicity. In
practice, it is also common to skip odd times relatively to a given initial time, or to
downsample using a higher factor than $2$. When using filtering and range restriction,
e.g., $\smash{C^{\downarrow}_{\lambda;t_i,t_f}}$, it is always understood that
downsampling is applied after restriction or filtering.

Downsampling is commonly used in lattice calculations when fitting correlators. Since
correlations are localized in time, discarding neighboring times naturally reduces
correlations. Since the Laplace filter also aims to cancel correlations in neighboring
time slices, an interesting question is whether its effect is compounded with or redundant
to a simple downsampling step.

We repeat the analysis of the previous section, but now for the downsampled correlators
$\smash{C^{\downarrow}_{\lambda}}$. The results are summarized in~\cref{fig:ds}. A first
observation is that, without any filtering applied, downsampling already reduces the total
CDR by $\unit{24}{\dB}$ and $\unit{19}{\dB}$ for M0M-mes-ll and F1M-mes-lh, respectively.
Optimizing the filter, as done in the previous section, leads to an additional decrease in
the total CDR by $\unit{41}{\dB}$ and $\unit{18}{\dB}$ for M0M-mes-ll and F1M-mes-lh,
respectively. On the restricted range, the CDR is improved by $\unit{41}{\dB}$ and
$\unit{21}{\dB}$ for M0M-mes-ll and F1M-mes-lh, respectively. These gains are similar to
those obtained without downsampling in the previous section, which suggests that Laplace
filtering improves correlations in a way that is not redundant with the additional gains
from downsampling. On the restricted range, the optimized CDRs for the downsampled data
are $\unit{11}{\dB}$ and $\unit{9}{\dB}$ for M0M-mes-ll and F1M-mes-lh, respectively. In
this configuration, the eigenvalues of the correlation matrix vary over roughly one order
of magnitude, ensuring that the ill-conditioning in computing the $\chi^2$ function
\cref{eq:chi2} has been largely eliminated.

We now summarize the various strategies explored for optimizing Laplace filters.
\subsection{Summary of correlation improvements}
\begin{figure*}[p]
    \includegraphics[width=.8\textwidth]{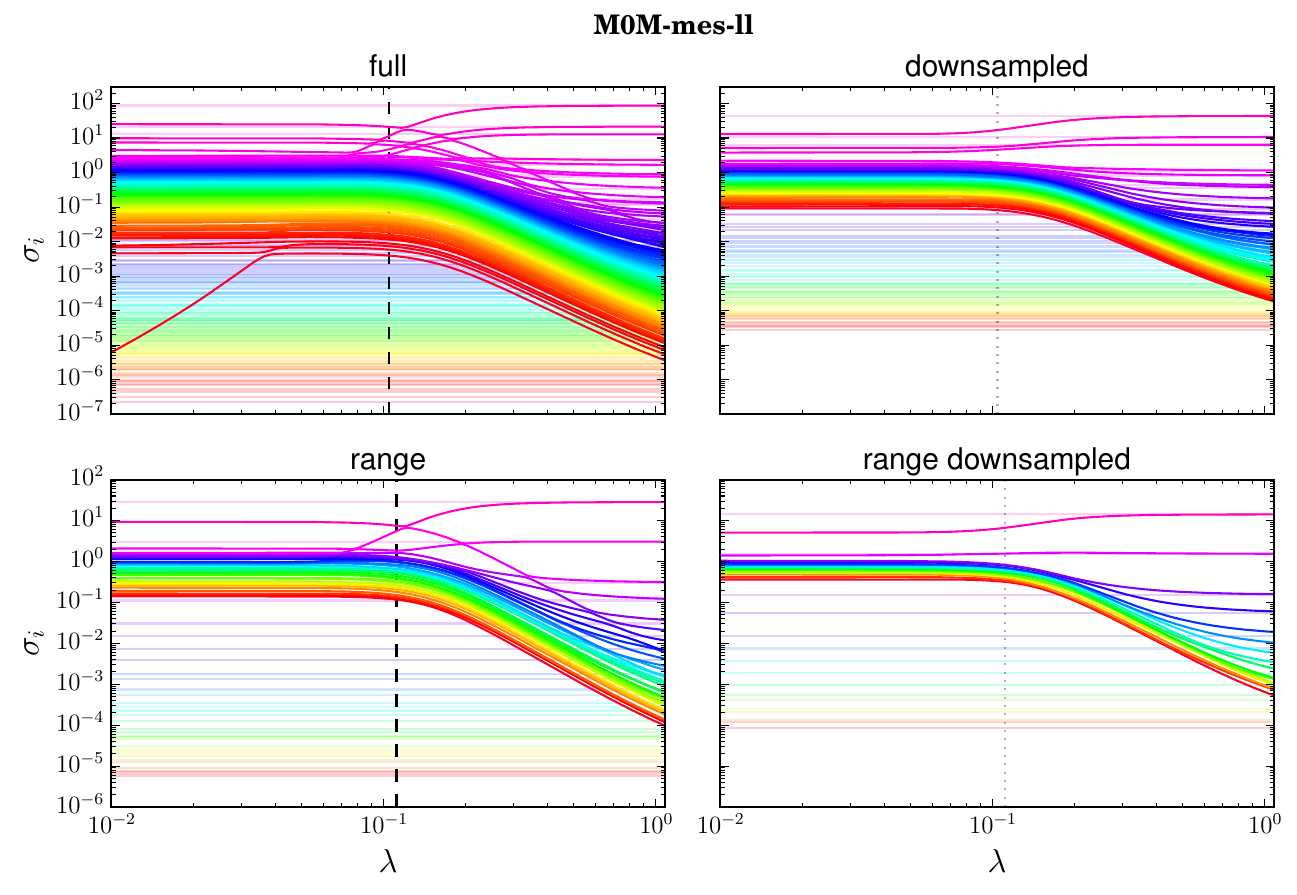}
    \includegraphics[width=.8\textwidth]{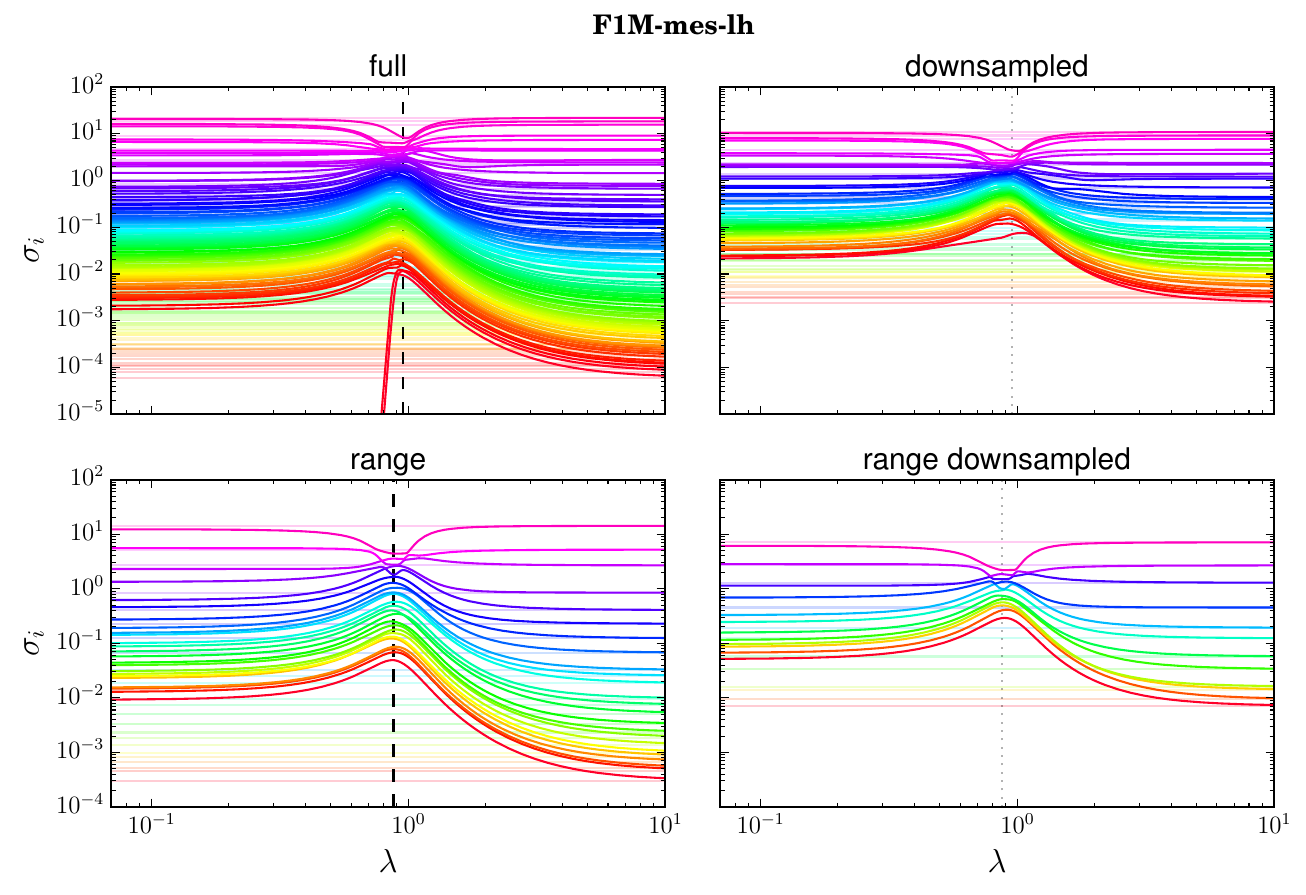}
    \caption{Behavior of the eigenvalues $\sigma_i$ of the correlation matrix of the 
    M0M-mes-ll (upper half) and F1M-mes-lh (lower half) correlation functions as a 
    function of Laplace filtering regulator $\lambda$. In each half, the top and bottom 
    rows show the spectrum for the full and restricted time ranges, respectively. In the 
    right column, an additional downsampling step is applied. The faint horizontal lines 
    correspond to the spectrum when no filtering is applied.}
    \label{fig:evs_vs_lambda}
\end{figure*}

 In~\cref{fig:evs_vs_lambda} we illustrate how the individual eigenvalues $\sigma_i$ of
the correlation matrix vary as a function of $\lambda$ for the M0M-mes-ll and F1M-mes-lh correlation functions.
The vertical dashed lines correspond to the values that minimise the CDR for the full
(top) or range-restricted correlation matrix. The faint horizontal lines correspond to the
unfiltered spectrum. We clearly observe the expected asymptotic behaviour of recovering
the original spectrum for very large values of $\lambda$ as well as the emergence of a
zero mode when considering the full correlation matrix and very small values of $\lambda$.
In~\cref{tab:cdr}, we summarize the CDRs obtained by filtering and downsampling the
M0M-mes-ll and F1M-mes-lh correlators, for both the full time range and a restricted,
representative fit range. In all cases, a combination of downsampling and optimized
filtering improves the condition number of the correlation matrix by at least four orders
of magnitude. This is an encouraging result, and we emphasize that this improvement is
achieved solely through a linear, invertible transformation of the data and does not
introduce any additional systematic error in subsequent fits of the correlators.

The filter optimizations discussed in this section only target improving correlations.
We have not discussed the effect of filtering on the statistical accuracy of the data,
which is critical for extracting physical information from the correlators. As observed
in~\cref{fig:optim,fig:ds}, many choices of $\lambda$ will lead to significant improvements in
correlations, which allows for the introduction of additional criteria related to the quality of
the filtered data. In the next section, we discuss the physical effect of the regulated
Laplace filter on lattice correlators and its impact on fitting spectral decomposition models to these data.

\section{Spectral effect of Laplace filters}
\label{sec:spectrum}
\begin{figure*}[p]
  \centering
  \includegraphics{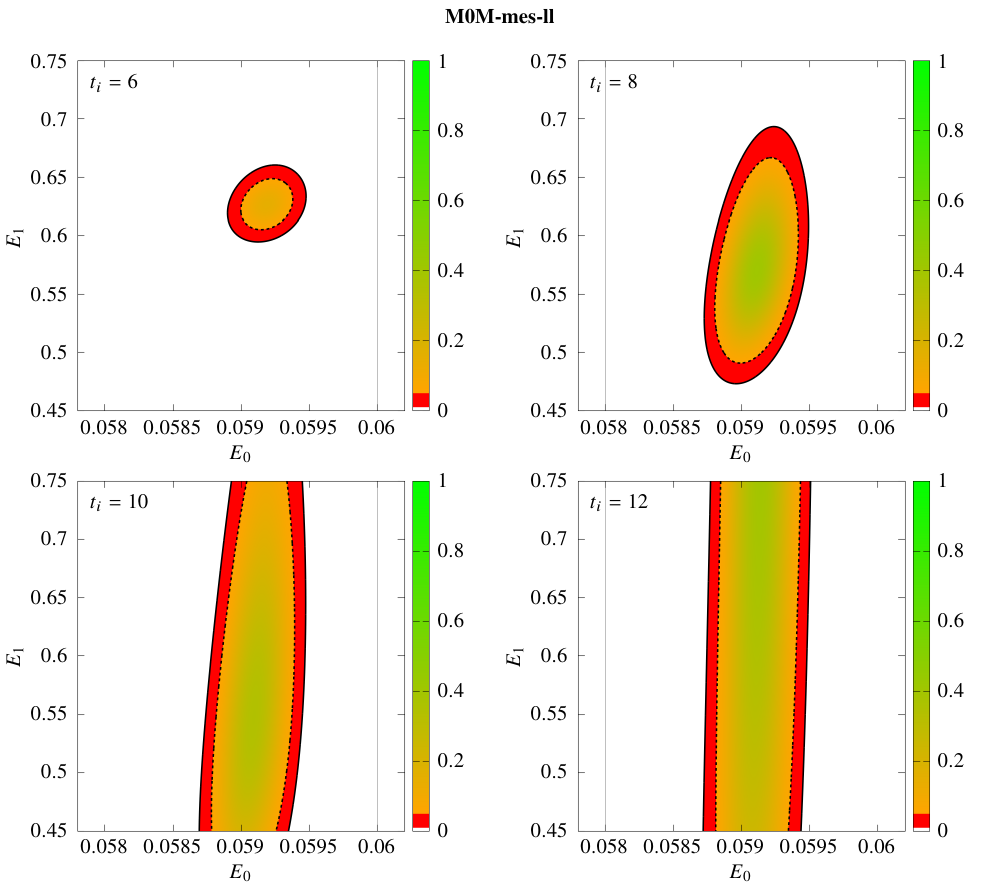}
  \caption{Representation of the pion correlator M0M-mes-ll 2-state test $p$-value
    $p_\Lambda$ defined in \cref{eq:plambda}, with $\Lambda=(\tilde{E}_0,\tilde{E}_1)$.
    The four different plots correspond to various values of the initial time $t_i$. In
    all cases, the final time is fixed to $t_f=48$. Values below $1\%$ are not represented
    (2-state hypothesis rejected at $99\%$ confidence level), and values above $1\%$ are
    within the solid black contour. Values between $1\%$ and $5\%$ are represented in red,
    and values above $5\%$ are within the dashed contour, and represented by a gradient
  between orange and green.}
  \label{fig:M0ll-scan-2d}
\end{figure*}
\begin{figure*}[p]
  \centering
  \includegraphics{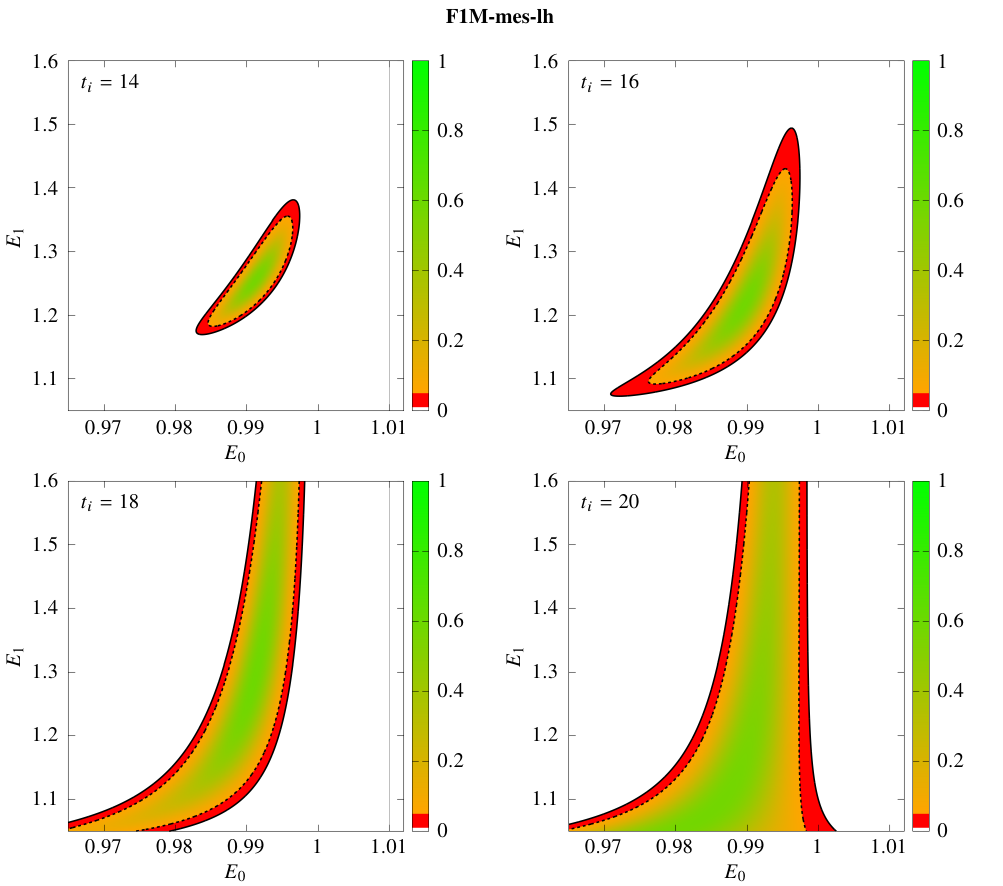}
  \caption{Representation of the heavy-light correlator F1M-mes-lh 2-state test $p$-value
    $p_\Lambda$ defined in \cref{eq:plambda}, with $\Lambda=(\tilde{E}_0,\tilde{E}_1)$.
    The four different plots correspond to various values of the initial time $t_i$. In
    all cases, the final time is fixed to $t_f=36$. Plotting conventions are identical to
  \cref{fig:M0ll-scan-2d}.}
  \label{fig:F1Mhl-scan-2d}
\end{figure*}

In this section, we discuss the effect of regulated Laplace filters on the spectrum of a
correlation function. We start with a theoretical discussion of this effect, and follow up
with applications to data analysis.
\subsection{Theoretical discussion}
\subsubsection{Continuous time}
In continuous Euclidean time, the Feynman Green's function for the propagation of a single
spin-0 state with nonzero mass $m$ and momentum $\mathbf{k}$ is given by
\begin{equation}
  G_0(t;E)=\int\frac{\diff k_0}{2\pi}\frac{e^{-i k_0t}}{k^2+m^2}
  =\frac{e^{-E|t|}}{2E}\,,
\end{equation}
where $k=(k_0,\mathbf{k})$ is the Euclidean four-momentum, and $E^2=\mathbf{k}^2+m^2$. By
construction, $G_0(t;E)$ is a solution to the differential equation
\begin{equation}
  (-\partial_t^2+E^2)\,G_0(t;E)=\delta(t)\,.
  \label{eq:green-fn-cont}
\end{equation}
A key observation here is that the left-hand side of the equation above is the continuous
version of the regulated Laplace filter \cref{eq:laplace} with $\lambda=E$. For an
arbitrary regulator value,
\begin{equation}
  (-\partial_t^2+\lambda^2)\,G_0(t;E)=\delta(t)+(\lambda^2-E^2)G_0(t;E)\,.
\end{equation}
In a field theory with a nonzero mass gap $M$, a general scalar two-point function $G(t)$
can be written for $t\neq 0$ using the Källén-Lehmann spectral representation
\begin{equation}
  G(t)=\int_M^{+\infty}\diff E\,\rho(E)G_0(t;E)\,,
  \label{eq:kl}
\end{equation}
where $\rho(E)$ is the associated spectral density. This is the standard result stating
that in Euclidean time, a two-point function is related to the Laplace transform of its
spectral density. Applying the Laplace filter to $G(t)$, one obtains
\begin{equation}
  (-\partial_t^2+\lambda^2)G(t)=\int_M^{+\infty}\diff E\,(\lambda^2-E^2)\rho(E)G_0(t;E)\,,
  \label{eq:filtered-kl}
\end{equation}
where the delta function term has been omitted since the spectral representation requires
$t\neq 0$.

In conclusion, from a spectral point of view, the Laplace filter with regulator $\lambda$
is equivalent to transforming the spectral density with $\rho(E)\mapsto
(\lambda^2-E^2)\rho(E)$. Let us now formulate the equivalent statement for discrete and
periodic time.
\subsubsection{Discrete and periodic time}
Making the lattice spacing explicit for this subsection, the Laplace filter defined in
\cref{eq:laplace}
in units of $a$ is given by
\begin{equation}
  \Delta_{\lambda} C[t]=a^{-2}[(2+a^2\lambda^2)\,C[t]-C[t-a]-C[t+a]]\,,
\end{equation}
which clearly has the continuum limit
\begin{equation}
  \lim_{a\to 0}\Delta_{\lambda}=-\partial_t^2+\lambda^2\,,
\end{equation}
when acting on samples of smooth functions. Then, one can show that the discrete version
of \cref{eq:green-fn-cont} is given by
\begin{equation}
  \Delta_{\tilde{E}}\,G_0[t;E]=a^{-1}\delta_{t,0}\,,
\end{equation}
where $\delta_{j,k}$ is the Kronecker delta, $G_0[t;E]$ is given by
\begin{equation}
  G_0[t;E]=\frac{e^{-E|t|}}{2a^{-1}\sinh(aE)}\,,
\end{equation}
and
\begin{equation}
  \tilde{E}^2=2a^{-2}[\cosh(aE)-1]\,.
\end{equation}
For $a\to 0$, $\tilde{E}$ converges to $E$ with leading $\mathcal{O}(a^2)$ corrections.
Switching back to lattice units (\ie, $a=1$), the equation above leads to the identity
\begin{equation}
  \Delta_\lambda(e^{-E|t|})=2\sinh(E)\,\delta_{t,0}+(\lambda^2-\tilde{E}^2)\,e^{-E|t|}\,,
\end{equation}
with $\tilde{E}^2=2[\cosh(E)-1]$ as before. We now consider a quantum field theory with a
nonzero mass gap $M$ and a discrete, periodic time dimension of length $N_t$. At nonzero
time separation $t$, an arbitrary scalar two-point function can be written as
\begin{equation}
  C[t]=\sum_nA_n\,g_0[t;E_n]\,,
  \label{eq:kllat}
\end{equation}
where $A_n$ and $E_n\geq M$ are finite sets of amplitudes and energies, respectively, and where
$g_0[t;E_k]$ is the periodization
\begin{equation}
  g_0[t;E]=\sum_{j=-\infty}^{+\infty}e^{-E|t+jN_t|}\,.
\end{equation}
\cref{eq:kllat} is the discrete and periodic equivalent of the K\"all\'en-Lehmann spectral
representation \cref{eq:kl}. For $t\in\{0,\dots,N_t-1\}$, the expression above can be reduced
to the more convenient form
\begin{equation}
  g_0[t;E]=\frac{e^{-Et}+e^{-E(N_t-t)}}{1-e^{-N_tE}}\,.
\end{equation}
For $t\neq 0\ (\mathrm{mod}\ N_t)$, we have
\begin{equation}
  \Delta_\lambda\,g_0[t;E]=(\lambda^2-\tilde{E}^2)\,g_0[t;E]\,,
\end{equation}
and therefore for $t\notin\{-1,0,1\}\ (\mathrm{mod}\ N_t)$
\begin{equation}
  \Delta_\lambda C[t]=\sum_n(\lambda^2-\tilde{E}_n^2)A_n\,g_0[t;E_n]\,,
  \label{eq:filtered-kllat}
\end{equation}
which is the discrete analog of \cref{eq:filtered-kl}. This last identity is only valid if
$t$ is at least two sites away from $0$; otherwise, the result will depend on the contact
term $C[0]$, which does not necessarily obey the spectral representation \cref{eq:kllat}.

An important conclusion from \cref{eq:filtered-kllat} is that the Laplace filter can
entirely remove the contribution of a given state from a correlator if the regulator is
tuned to the energy $\tilde{E}$ associated with this state. On the one hand, this means that
when choosing the regulator, one must be careful to avoid the energies of states of
interest. On the other hand, it also implies that the filter can be exploited to remove
unwanted states, typically excited-state contamination from hadronic interpolating
operators.
\subsubsection{Summary and generalization}
A key ingredient of the discussion above is the identity
\begin{equation}
  \Delta_\lambda(e^{\pm Et})=(\lambda^2-\tilde{E}^2)\,e^{\pm Et}\,.
\end{equation}
Euclidean time correlators are, however, only linear combinations of such exponentials
when all pairs of operators involved have nonzero time separation. This is a particularly
important constraint with discrete time, since the filtered correlator
$\Delta_{\lambda}C[t]$ depends on the times $C[t-1]$ and $C[t+1]$. If these neighboring
terms are not contact terms, then the Laplace filter is a spectral reweighting according
to the equation above. When time is additionally periodic of size $N_t$, the notion of
nonzero time separation must be understood modulo $N_t$.

These general principles also apply to $n$-point functions, as well as to correlation
functions involving nonzero-spin states. Indeed, an arbitrary $n$-point time correlator
will depend on $n-1$ time separations and can be written as a sum of exponentials if all
time separations are nonzero (mod $N_t$). Additionally, nonzero-spin states also propagate
with an exponential time dependence of the form $e^{\pm Et}$. The coefficients of such
exponentials will be spin matrices and will generally differ between the backward and
forward time directions, however properties of the Laplace filters discussed in this paper
can be generalized to these cases.
\subsection{Applications to statistical data analysis}
Throughout this section, we assume that the correlator under study is given by
\cref{eq:kllat}, with $N_s$ states of nonzero amplitude and distinct energies indexed by
$n\in\{0,\dots,N_s-1\}$. For $t\notin\{-1,0,1\}\ (\mathrm{mod}\ N_t)$, a single application
of the Laplace filter yields \cref{eq:filtered-kllat}. For a vector of $r$ regulators
$\Lambda=(\lambda_0,\dots,\lambda_{r-1})$, we define
\begin{equation}
  C_{\Lambda}[t]=\Delta_{\lambda_1}\cdots\Delta_{\lambda_r} C[t]\,,
\end{equation}
i.e., $\smash{C_{\Lambda}}$ is the correlator transformed by $r$ successive Laplace
filters. \cref{eq:filtered-kllat} can be generalized to multiple filters. For
$t\notin\{-r,\dots,r\}\ (\mathrm{mod}\ N_t)$, one obtains
\begin{equation}
  C_{\Lambda}[t]=
  \sum_n\left[\prod_{j=0}^{r-1}(\lambda_j^2-\tilde{E}_n^2)\right]A_n\,g_0[t;E_n]\,.
  \label{eq:nfilter-kllat}
\end{equation}
For the remainder of this discussion, we assume that $t\notin\{-r,\dots,r\}\
(\mathrm{mod}\ N_t)$. In order for the formula above to be valid for at least one
time slice, we require $r<N_t/2$. Then, \cref{eq:nfilter-kllat} offers an opportunity to
determine the energy spectrum $\smash{\tilde{E}_n}$ by finding a sequence of Laplace
filters that annihilate $C[t]$, i.e., that solve the equation
\begin{equation}
  C_{\Lambda}[t]=0\,,  \label{eq:nfilter-eq}
\end{equation}
for all $t$ in a given range satisfying the previous constraints, and containing at least
$r$ elements. If the number of filters equals the number of states, i.e., $r=N_s$, then
\cref{eq:nfilter-eq} is satisfied if and only if, for all $j\in\{0,\dots,N_s-1\}$,
\begin{equation}
  \lambda_j = \tilde{E}_j\,.
\end{equation}
In this case, the energies $E_n$ are uniquely determined\footnote{This is a
  consequence of the fact that functions of the form $t\mapsto e^{-Et}$ are linearly
independent for distinct values of $E$.} by \cref{eq:nfilter-eq}. If $r>N_s$, then necessarily the spectrum is a subset of the solution set of
\cref{eq:nfilter-eq}, \ie,
$\smash{\{\tilde{E}_0,\dots,\tilde{E}_{N_s-1}\}\subset\{\lambda_0,\dots,\lambda_{r-1}\}}$. If
$r<N_s$, then \cref{eq:nfilter-eq} has no solution.

In summary, if $N_s<N_t/2$, the spectrum of $C$ can be determined as follows:
\begin{enumerate}
  \item The number of states $N_s$ is given by the minimal $r$ for which
    \cref{eq:nfilter-eq} admits a solution;
  \item The spectrum $\tilde{E}_j$ is given by the unique solution of \cref{eq:nfilter-eq}
    for $r=N_s$.
\end{enumerate}

The procedure above can serve as the basis for new methods to determine the energy
spectrum of lattice correlators. Although equation~\cref{eq:nfilter-eq} determines the
spectrum uniquely, in practice with noisy simulation data it can only be solved up to
statistical noise, introducing ambiguities in the above procedure. Therefore, 
in what follows, we discuss potential hypothesis-testing methods to constrain the 
spectrum of the correlator.
\subsubsection{Multistate hypothesis testing}
\begin{figure*}[t]
  \centering
  \includegraphics{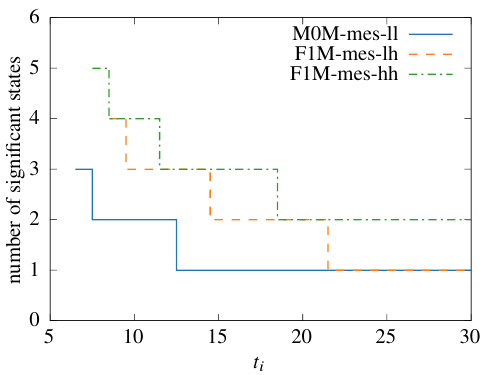}
  \includegraphics{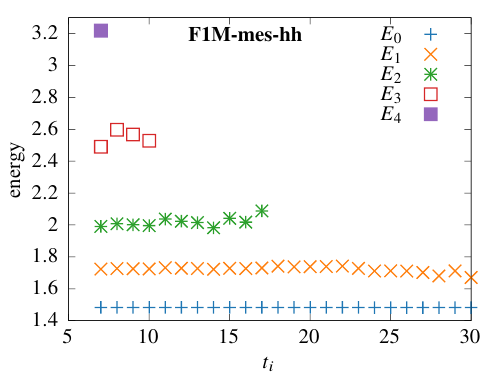}
  \caption{The left plot represents the number of significant states as a function of the
    initial time $t_i$, using the procedure described in \cref{sec:sig-states}, at the
    $95\%$ confidence level, for the pion (M0M-mes-ll), heavy-light (F1M-mes-lh), and
    heavy-heavy (F1M-mes-hh) pseudoscalar two-point functions. For M0M-mes-ll, the final
    time is fixed to $t_f=48$, and $t_f=36$ in the two other cases. If the procedure is
    inconclusive (\ie~$H_s$ was rejected at step 3), no points are plotted for the
    associated values of $t_i$. The right plot represents the energies of the significant
    states of the F1M-mes-hh correlator, obtained as a side product of the analysis
  performed in the left plot.}
  \label{fig:sig-states}
\end{figure*}
A key observation in the discussion above is that \cref{eq:nfilter-eq} has no solution if
the number of filters is smaller than the number of states, \ie, $r<N_s$. This statement
can be reformulated statistically as a hypothesis rejection, which allows us to design
tests for the number of statistically significant states in a given time range.

We consider a time range $\{t_i,\dots,t_f-1\}$ for which the condition
$t\notin\{-r,\dots,r\}\ (\mathrm{mod}\ N_t)$ is always satisfied. For sufficiently large
statistics, the average restricted correlator $\bar{C}_{t_i,t_f}$ is distributed according
to a normal distribution $\mathcal{N}(C_{t_i,t_f},\Sigma_{t_i,t_f})$ with mean
$\bar{C}_{t_i,t_f}$ and covariance $\Sigma_{t_i,t_f}$, where $\Sigma$ is the true covariance
matrix estimated in practice by the matrix $V$ defined in \cref{eq:covmat}. Similarly to
\cref{eq:restr2}, we define
\begin{equation}
  C_{\Lambda;t_i,t_f}=R_{t_i,t_f}C_{\Lambda}\,,
\end{equation}
and we denote $\Sigma_{\Lambda;t_i,t_f}$ as its covariance matrix. Since the Laplace
filter is linear, we know that $\bar{C}_{\Lambda;t_i,t_f}$ is distributed according to
$\mathcal{N}(C_{\Lambda;t_i,t_f},\Sigma_{\Lambda;t_i,t_f})$. For a given $\Lambda$, we can
consider the null hypothesis $H_\Lambda$ defined by
\begin{equation}
  C_{\Lambda;t_i,t_f}=0\tag{$H_\Lambda$}\,.
\end{equation}
For a given significance level $\alpha$, rejecting $H_\Lambda$ means that there exists at
least one element of the spectrum of $C$ which is not part of $\Lambda$, with a confidence
level of $1-\alpha$. Since $C_{\Lambda;t_i,t_f}$ is normally distributed, such a test can
be performed using the usual $T^2$ statistic
\begin{equation}
  T^2(\Lambda)=
  \bar{C}_{\Lambda;t_i,t_f}^{T}V_{\Lambda;t_i,t_f}^{-1}\bar{C}_{\Lambda;t_i,t_f}\,,
  \label{eq:t2}
\end{equation}
which is asymptotically distributed according to a $\chi^2$ distribution with $\Delta t=t_f-t_i$ degrees
of freedom. The $p$-value of the test is then given by the probability
\begin{equation}
  p_{\Lambda}=\Prob[T^2(\Lambda)>T^2_{\mathrm{obs}}(\Lambda)]=1-F[T^2_{\mathrm{obs}}(\Lambda),
  \Delta t]\,,
  \label{eq:plambda}
\end{equation}
where $T^2_{\mathrm{obs}}(\Lambda)$ is the $T^2$ statistic observed from the data, and
$F(x,\nu)$ is the cumulative distribution function of the $\chi^2$ distribution with $\nu$
degrees of freedom. Finally, if $p_\Lambda<\alpha$, then $H_{\Lambda}$ is
rejected. In the remainder of this section, we will use $\alpha=0.05$.

As an example, we show in \cref{fig:M0ll-scan-2d,fig:F1Mhl-scan-2d} representations of
$p_\Lambda$ for two states $\Lambda=(\tilde{E}_0,\tilde{E}_1)$, and for various values of
the initial time $t_i$. At early values of $t_i$, we observe that $H_{\Lambda}$ is
systematically rejected except for a well-localized $p_\Lambda>\alpha$ region in the
$(E_0,E_1)$ plane. In this region, the data can be described up to noise as the sum of two
exponentials with well-constrained energies. As the initial time increases, we observe
that the $p_\Lambda>\alpha$ region eventually does not constrain the location of $E_1$
anymore. This suggests, as expected, that for sufficiently large times the correlator can
be statistically represented by a single exponential.

The $p$-value $p_{\Lambda}$ allows us to potentially reject $\Lambda$ as a statistically
acceptable spectrum to describe the correlator. However, this is only done at a specific
value of $\Lambda$, and it is interesting to design a global test for multistate
hypotheses. We start by defining the following $r$-state null hypothesis
\begin{equation}
  \text{there exists }\Lambda=(\lambda_0,\dots,\lambda_{r-1})\text{ s.t. }
  C_{\Lambda;t_i,t_f}=0\tag{$H_r$}\,.
\end{equation}
$H_r$ can be tested by minimizing the statistic $T^2(\Lambda)$ over
$\Lambda=(\lambda_0,\dots,\lambda_{r-1})$:
\begin{equation}
  T^2_r=\min_{\lambda_0,\dots,\lambda_{r-1}}T^2(\lambda_0,\dots,\lambda_{r-1})\,.
\end{equation}
Similarly to curve fitting, such a minimum requires the gradient of
$T^2(\lambda_0,\dots,\lambda_{r-1})$ to vanish. This constraint makes $T^2_r$ a sum of $\Delta
t-r$ independent Gaussian random variables squared, and therefore $T^2_r$ asymptotically obeys the
$\chi^2$ distribution with $\Delta t-r$ degrees of freedom. We therefore define the
$r$-state $p$-value as
\begin{equation}
  p_r=\Prob(T^2_r>T_{r,\mathrm{obs}}^2)=1-F(T_{r,\mathrm{obs}}^2,\Delta t-r)\,,
\end{equation}
where $T_{r,\mathrm{obs}}^2$ is the minimized $r$-state $T^2$ obtained from the data. In
practice, such a minimum can be computed using standard minimization algorithms with
\cref{eq:t2}. If one rejects $H_r$ with $p_r<\alpha$, it means that with $1-\alpha$
confidence level, the data cannot be described as a sum of $r$ exponentials on the chosen
time range. However, if $H_r$ is not rejected, it does not necessarily imply that all $r$
exponentials contribute significantly to the correlator -- it is possible that a subset of
$\Lambda$ is sufficient. We now discuss an extended procedure to address this.
\subsubsection{Number of significant states}
\label{sec:sig-states}
For a hypothesized number of states $r$, we want to determine whether a $(r+1)$-state
description of $C$ over the time range $\{t_i,\dots,t_f-1\}$ would provide a significant
improvement over the hypothesis. This can be done using a standard likelihood-ratio test,
considering $H_{r}$ as the null hypothesis and $H_{r+1}$ as the alternative hypothesis. To
achieve this, one first computes the difference in the minimum $T^2$ statistic between the
two models, \ie,
\begin{equation}
  \Delta T^2_r=T^2_{r+1}-T^2_{r}\,.
\end{equation}
Wilks' theorem~\citep{Wilks:1938dza} then implies that $\Delta T^2_r$ is asymptotically distributed as a
$\chi^2$ distribution with one degree of freedom, and we therefore define the
corresponding $p$-value
\begin{equation}
  \bar{p}_r=\Prob(\Delta T^2_r>\Delta T_{r,\mathrm{obs}}^2)=1-F(\Delta
  T_{r,\mathrm{obs}}^2,1)\,,
\end{equation}
where $\Delta T_{r,\mathrm{obs}}^2$ is the observed value of $\Delta T^2_r$ from the data.
If $\bar{p}_r<\alpha$, then $H_{r}$ is \emph{rejected in favor of} $H_{r+1}$; that is, the
$(r+1)$-th state is a statistically significant addition to the description of the data.
As is usual with ratio tests, it is important to note that only relative significance is
tested, which is independent of whether $H_r$ and $H_{r+1}$ can be individually rejected.
The values $\bar{p}_r$ allow one to determine the \emph{number of significant states} in a
chosen time range, which we propose to do using the following procedure.
\begin{enumerate}
  \item Choose a maximum number $r_{\mathrm{max}}>1$ of states to test, and compute the
    values $p_r$ for $r\in\{1,\dots,r_{\mathrm{max}}\}$ and $\bar{p}_r$ for
    $r\in\{1,\dots,r_{\mathrm{max}}-1\}$.
  \item Using the $\bar{p}_r$ values, find the maximum value $s$ for which $H_{s-1}$ is
    rejected in favor of $H_{s}$. If no such rejection occurs, then $s=1$.
  \item If $p_{s}$ does not result in a rejection of $H_{s}$, then the number of
    significant states is $s$ at the $1-\alpha$ confidence level.
\end{enumerate}
In step 1, it is important to note that the constraint $t_i>r_{\mathrm{max}}$ must hold to
satisfy the requirements of \cref{eq:nfilter-kllat}. In step 2, one must be careful that
as $r_{\mathrm{max}}$ increases, naively conducting multiple tests of the form $\bar{p}_r<\alpha$ 
will increase the probability of erroneous rejections, effectively lowering the overall confidence level.
This is a well-known issue in statistics generally referred to as \emph{family-wise error rate}. A confidence level of
$1-\alpha$ can be enforced using the
Holm-Bonferroni~\citep{holmSimpleSequentiallyRejective1979} method, which lowers the rejection thresholds to
\begin{equation}
  \bar{p}_{r_j}<\frac{\alpha}{r_{\mathrm{max}}+1-j}\,,
\end{equation}
where $j\in\{1,\dots,r_{\mathrm{max}}\}$ is such that the sequence $\bar{p}_{r_j}$ is ordered
from the lowest to the highest $p$-value.
It is important to notice that family-wise error
rate control methods generally require deciding in advance the number of hypothesis to be
tested, which is part of the justification for step 1. Finally, if $H_s$ is rejected in
step 3, we consider the procedure to be inconclusive. This can happen, for example, if
$r_{\mathrm{max}}$ is too small and not enough states were added to obtain an acceptable
description of the data.

In \cref{fig:sig-states}, we show the result of applying the procedure above to
M0M-mes-ll, F1M-mes-lh, and F1M-mes-hh. The outcome is as expected: the number of
significant states decreases as the initial time increases. Charmonium-like correlators
are well known to generally suffer from considerable excited-state contamination, and we
indeed observe from our representative data (F1M-mes-hh) that there are no initial times
compatible with a single-state description. We find the procedure to be particularly
promising for determining safe minimum times to attempt a single-state fit to the data.
Indeed, if the number of significant states is equal to $1$, then the procedure above
guarantees that the data are compatible with a single-state description at the $95\%$
confidence level. In the data presented in \cref{fig:sig-states}, the earliest safe
initial time for single-state fits is $t_i=12$ for M0M-mes-ll, and $t_i=21$ for
F1M-mes-lh. It is crucial to note that the proposed procedure is able to establish these
results \emph{with no knowledge of the amplitudes $A_n$}, which makes it more
attractive than attempting a non-linear multi-exponential fit test on the data.
\subsubsection{Possible extensions}
As part of the procedure proposed above, one also obtains the energy values for the
significant states through the minimization of $T^2(\Lambda)$, opening the way for novel
methods to determine energies from lattice data, without requiring multi-exponential model
fits to the data. For example, we show in the right panel of~\cref{fig:sig-states} the
energies of the significant states for the F1M-mes-hh obtained through the analysis
described in the previous section. The values obtained are remarkably stable, and an
encouraging sign that determining the spectrum through the response of applying Laplace
filters is feasible. We will explore this further in future publications. In particular,
one needs to understand how to rigorously define the statistical uncertainties of the
energies obtained in~\cref{fig:sig-states}. Additionally, this new method essentially
determines the energies as the coefficients of a finite-difference equation canceling the
correlator. This is also the mathematical foundation of the Prony method, and comparison
with the outcome of solving a Prony GEVP is an interesting extension. However, due to the
iterative nature of our method, we can already state that our method does not suffer from
issues or ambiguities related to ordering eigenvalues, which is a known problem with
GEVP-based methods.

Finally, in all this section, we based hypothesis testing on asymptotic distributions,
assuming a large statistics of independent normally distributed samples. In practice the
exact probability distributions of the various estimators used will be affected by the
unavoidably finite statistics used in data analysis, as well as the fact that typical MCMC
samplers used in lattice QCD produce autocorrelated sequences. We expect that the method
presented here can be generalized to using empirical distributions, for example obtained
using bootstrap resampling. Such approach is discussed in \citep{Christ:2024nxz} for
standard correlator analysis, and we expect it can be extended to the methodology we
discussed in this section.

We now discuss applying the various methods in this paper to realistic data analysis for the extraction of energies and matrix elements from our representative dataset.

\section{Application examples}
\label{sec:application}
In this section we
discuss how the application of Laplace filtering can be used to improve fits to real
lattice data. Lattice correlation functions are generally fit to a model of the form
\begin{equation}
  C[t] = \sum_{n=0}^{N-1} A_n e^{-E_nt}\,,
  \label{eq:specdecomp}
\end{equation}
where the amplitudes $A_n$ and energies $E_n$ are model parameters. As discussed in the
previous section, the time dependence of an arbitrary $n$-point function is generally more
complicated; however, it is common practice to analyze such functions according to a
single time separation, and in an asymptotic time regime where the model above is a
suitable description, up to negligible exponential corrections coming from higher-energy
states and periodic contributions in time. Energies and amplitudes are then extracted by
means of a $\chi^2$ minimization, \ie~by fitting some restricted time range of the data to
\cref{eq:specdecomp} for a finite number of states $N$. As outlined in
\cref{sec:data-trans}, this requires the inversion of the estimated correlation matrix
between the data points that enter the fit. As discussed in \cref{sec:laplace}, applying Laplace
filtering allows one to find a representation of the data that significantly reduces its
CDR, making the numerical inversion of the correlation matrix more reliable. However, to
be of practical use, we also require that the statistical uncertainties of extracted
observables such as energies and amplitudes remain under control. Anticipating the results from~\cref{sec:mesfits}, we find that for the
datasets under consideration here, the value of $\lambda_0$ that minimizes the CDR leads
to larger statistical uncertainties on the extracted parameters $A_n$ and $E_n$. However,
as can be inferred from \cref{fig:optim,fig:ds}, there is a wide range of values of
$\lambda$ around $\lambda_0$ for which the CDR is reduced. In fact, based on the
discussion about spectral aspects in the previous section, we will see that in practice
there are additional benefits of Laplace filtering that can be exploited.

Most correlation functions in lattice QCD face a signal-to-noise problem as a function of
Euclidean lattice time $t$, where the signal decays faster than the square root of the
covariance, rendering earlier time slices statistically more precise than later ones. On
the other hand, one is commonly interested in the lowest-lying state of the
spectrum with energy $E_0$. So any optimal fit is necessarily a trade-off between
leveraging early time slices and controlling their contamination by states $n\geq N$ that
are not captured in the ansatz. Recalling the effect of Laplace filtering on the
correlation function, namely
\begin{equation}
  D_\lambda C[t]
  = \sum_{n=0}^{N-1} A_n(\lambda^2-\tilde{E}_n^2)\,e^{-E_nt}
  = \sum_{n=0}^{N-1} A_{\lambda;n}\,e^{-E_nt}\,,
  \label{eq:2ptfiltered}
\end{equation}
where $\tilde{E}_n^2=2[\cosh(E_n)-1]$, we observe that the parameter $\lambda$ can be
chosen to eliminate (or at least significantly reduce) a particular state $n$ by choosing
$\smash{\lambda=\tilde{E}_n}$.

\begin{figure*}[p]
  \includegraphics[width=\textwidth]{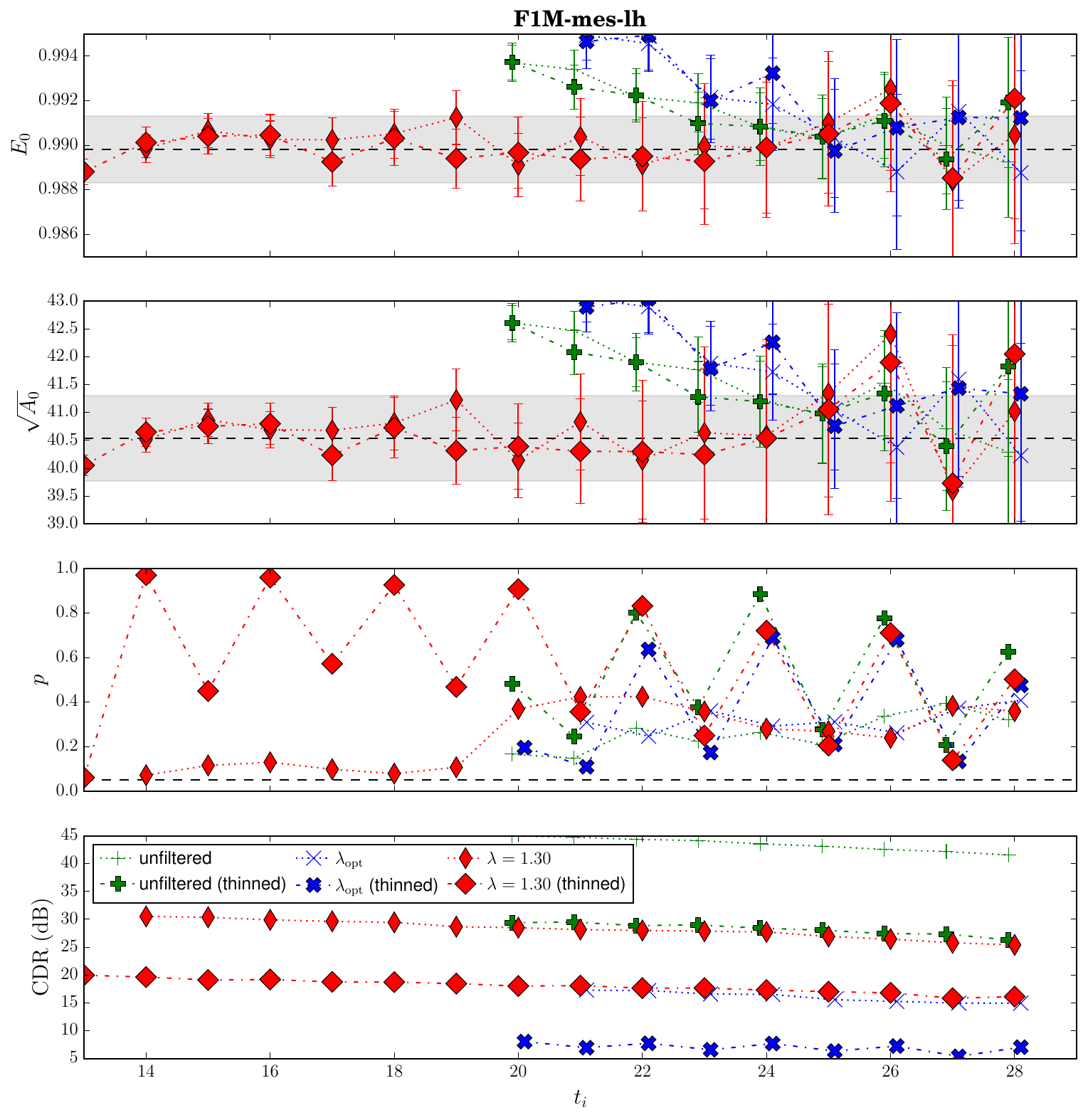}

  \caption{Fit results for single exponential fits to the F1M-mes-lh dataset as a
    function of $t_i$ with fixed $t_f = 3/8 N_t$. The green pluses,
    blue crosses and red diamonds represent the fit results to the unfiltered data, the
    filtered data for which the CDR of the entering time slices is minimized, and the
    choice $\lambda = 1.3$, respectively. In each case the smaller symbols represent fits
    to the entire indicated time range, while the larger symbols include an additional
    down-sampling step. The gray horizontal bands in the top two panels show results
    obtained from a stability analysis of two-exponential fits to multiple 2 and 3 point
    functions and serve as validation of the obtained fit results.} \label{fig:mes}
\end{figure*}

Systems involving a light $(u,d,s)$ and a heavy $(b,c)$ quark are particularly prone to
excited-state contamination due to their signal-to-noise properties and the relatively
small gap between the ground state and the lowest excited states of a system. In the
following, we will therefore illustrate the potential of our method on the F1M-mes-lh,
C1M-semi-sc-ss, and F1M-mix-sh datasets.

\subsection{Mesonic 2-point functions\label{sec:mesfits}}
In \cref{fig:mes} we show fit results obtained from a simple single-state (\ie~$N=1$ in
\cref{eq:specdecomp}) fit to the F1M-mes-lh dataset. We hold the maximum fit time
\mbox{$t_f=3N_t/8 = 36$} fixed and vary $t_i$ (shown on the $x$-axis).
Making use of the symmetry present in this particular correlation function, all fits were
performed to \emph{folded} data, meaning that the correlation function has been symmetrized 
for time reversal (\ie~$C[t]\mapsto(C[t]
+ C[N_t-t])/2$, with the restriction $t \leq N_t/2$) prior to any subsequent
operation. This is of practical convenience but not a necessity for the suggested method
to work.

Only fits with an acceptable ($\geq 5\%$) $p$-value are shown. The top panel shows the fit
results for the ground state energy, the second panel the value $\sqrt{A_n}$. The third
panel reports the $p$-value associated with this fit and the last the CDR of the
correlation matrix that is inverted.
Three different types of fits are shown; in each case, the fit is performed once
to the full time range (smaller symbols) and once to the down-sampled or thinned version
where only every second time slice enters (larger symbols).

The first type (green pluses) are fits to the unmodified data. We observe that the
earliest fits with an acceptable $p$-value occur around $t_i\approx 20$, but
the central values still shift until they reach a stable value around
$t_i\approx 24$. This is independent of whether the fit is performed to the
full or the thinned time range. We note that this is consistent with the determination of the number of significant states presented in~\cref{fig:sig-states} which found that the original correlation function can be described by a single state from $t_i \geq 21$.

The second type of fit (blue crosses) is performed by
first determining the value of $\lambda_\mathrm{opt}$ that minimizes the CDR of the data
entering the fit. Performing the fits to the data filtered by this value, we again observe
that the earliest $t_i$ yielding acceptable fits is $t_i \approx 20$
and even larger values of $t_i$ are required in order to reach stable fit
results. Furthermore, the statistical uncertainties on these fit results are larger than
their unfiltered counterparts. Unfortunately, this implies that the value which minimizes the
correlations does not provide a better fit result. However, as can be seen from 
\eg~\cref{fig:optim}, there typically is a region with $\lambda > \lambda_\mathrm{opt}$ that
still reduces the CDR. Furthermore, it is expected that there is a range where
cancellations of the excited-state contributions occur in the correlation function. For
the correlation function at hand, we find that a value of $\lambda = 1.3$ (third fit type;
red diamonds) achieves both: The CDR is significantly reduced compared to the unfiltered
data set (see bottom panel). At the same time, fit results with acceptable $p$-values are
already achieved for values $t_i \approx 13$ and stabilize from $t_i
\approx 14$. In addition to these benefits, the statistical uncertainty of the fit result
is notably reduced compared to the first stable values of the unfiltered or the `optimal'
value.  This is a promising result: Laplace filtering the data with a suitable choice of
filter allows one to reduce the contamination of excited states at comparably early time
slices, thereby exploiting statistically more precise data points. These findings are not
restricted to the energy of the state (top panel) but apply also to the amplitude, which
in many practical applications is the main quantity of interest.

For all three types of fits, when additionally applying thinning (larger symbols), the CDR
is further reduced and the $p$-values of the resulting fits are typically quantitatively
better, but the fit results remain largely unchanged.
For convenience, results from an extensive stability analysis
  entirely analogous to the one presented in Ref.~\citep{Boyle:2024gge} are
  superimposed and shown by the gray bands in the top two panels
  of~\cref{fig:mes}. The choice of $\lambda = 1.30$ was initially found by
  tuning $\lambda$ with the goal of lowering the $t_i$ for which
  acceptable fits first occur. However, we can interpret this value by noting
  that the stability analysis found $E_1 = 1.222(27)$, hence resulting in
  $\tilde{E_1} = 1.29946$. Hence filtering with $\lambda = 1.30$ effectively
  eliminates the first excited state from the correlation function. Furthermore,
  from~\cref{fig:sig-states} we then expect that after removing the first
  excited state, the filtered correlation function can be described by a single
  state from $t_i\approx 15$ in agreement with our findings.

\begin{figure*}
  \includegraphics[width=\columnwidth]{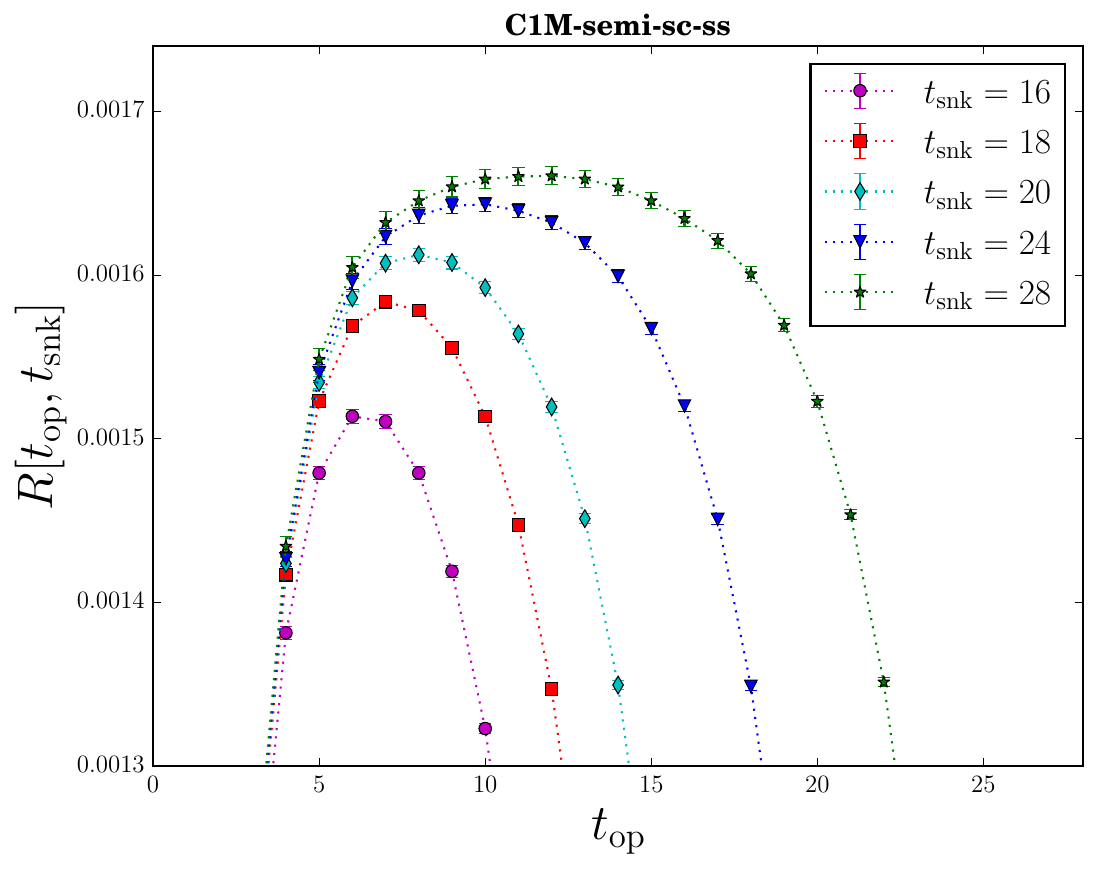}
  \includegraphics[width=\columnwidth]{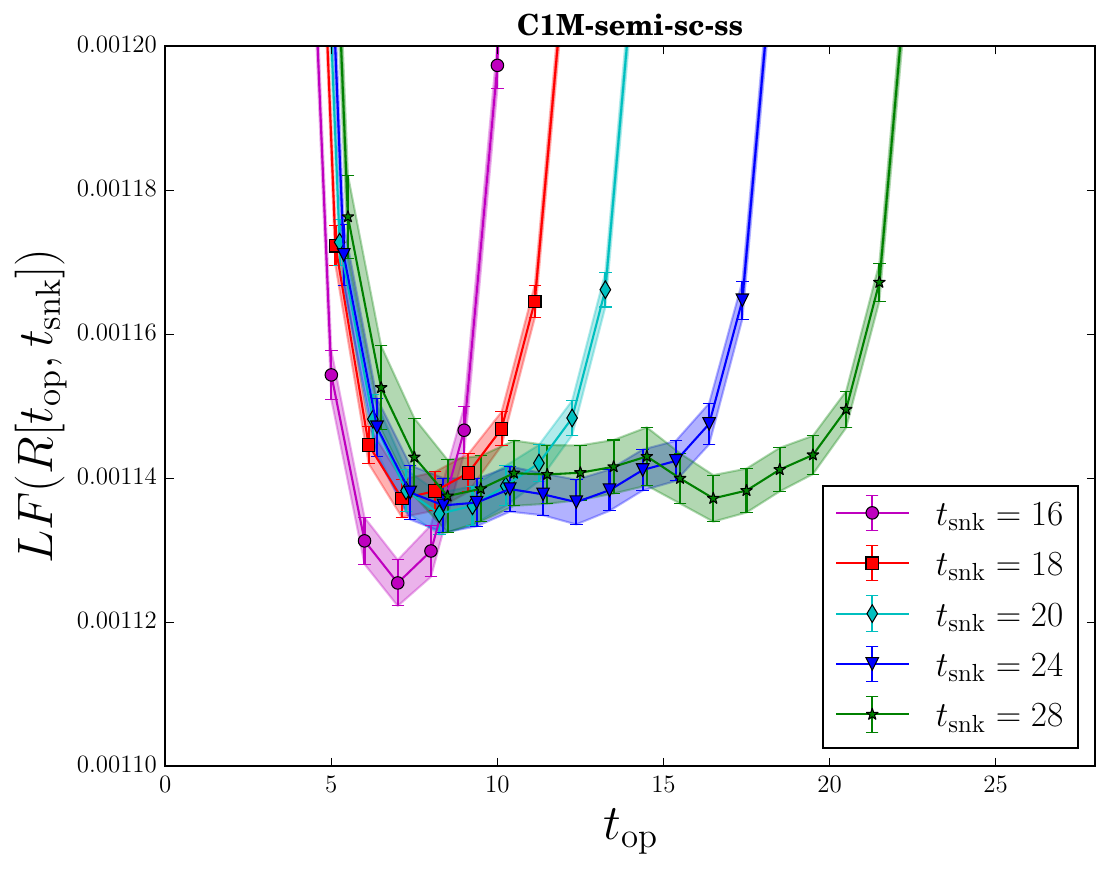}
  \caption{Illustration on the C1M-semi-sc-ss dataset of ratios that for sufficiently
    large times $t_\mathrm{op}$ and $t_\mathrm{snk}$ should approach an asymptotic value.
    The left-hand panel shows the unfiltered data, where no region that is independent of
    $t_\mathrm{snk}$ can be isolated. The right panel shows the data after
    Laplace filtering
    the two point functions that enter this ratio. Here a region for which
    three different
    choices of $t_\mathrm{snk}$ coincide for values $0 \ll t_\mathrm{op} \ll
    t_\mathrm{snk}$
  can be identified.} \label{fig:semilep}
\end{figure*}

\begin{figure}
  \includegraphics[width=\columnwidth]{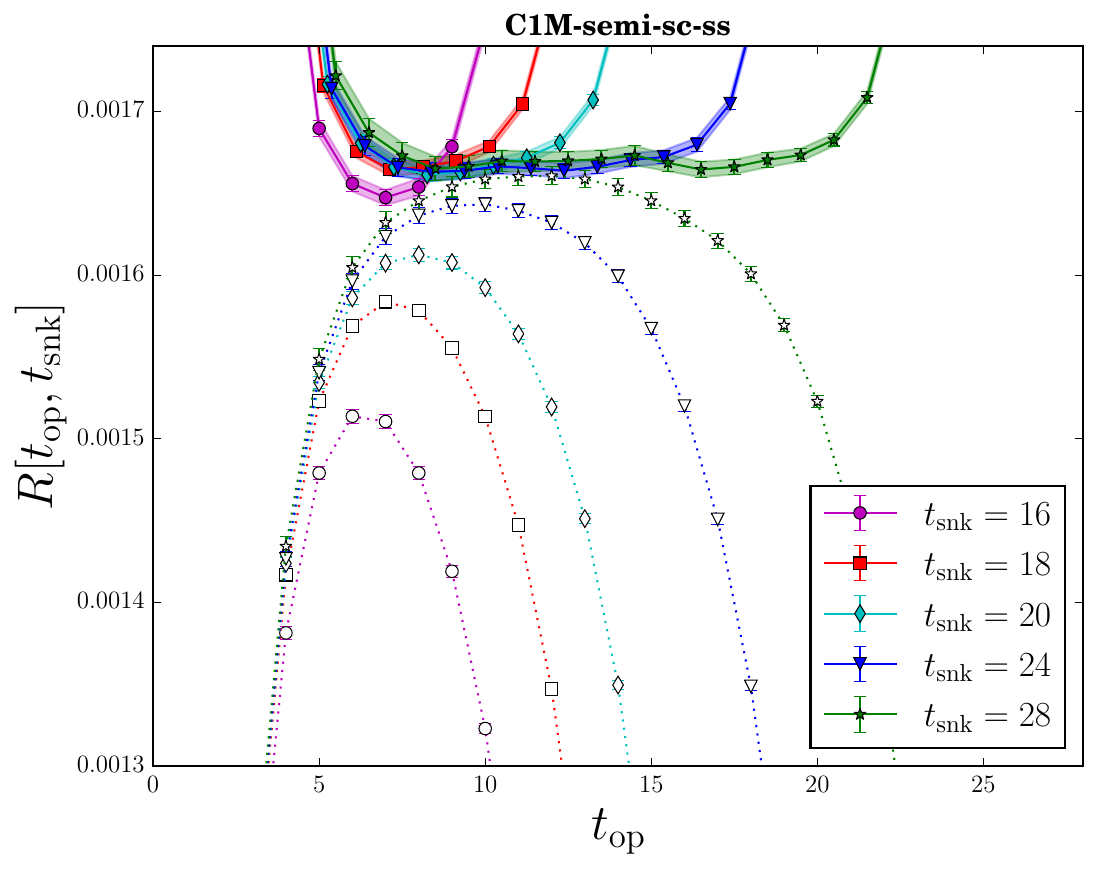}
  \caption{The same data as in \cref{fig:semilep}. The unfiltered data is shown with open
    symbols, whilst the filtered data has been rescaled by the appropriate factor of
    $(\lambda_{D_s}^2-(\tilde{E}_0^{D_s})^2)(\lambda_{\eta_s}^2-(\tilde{E}_0^{\eta_s})^2)$,
  assuming $E_0^{D_s} = 1.154$ and $E_0^{\eta_s} = 0.4030$.}
  \label{fig:semilep_samescale}
\end{figure}

\subsection{Application to semileptonic decays}
\label{subsec:semi}
In the following, we will explore the application of our method to more complicated
processes. We start by considering the semileptonic decay of a $D_s$ meson to an $\eta_s$
meson, where a charm quark decays to a strange quark. For demonstration, we only consider
the zero recoil (also known as $q^2_\mathrm{max}$) data point, where the initial and the
final hadronic states are at rest. In addition to the $\eta_s$ and $D_s$ 2-point functions
(see \cref{eq:specdecomp}), the three point function $C^{cs\to
ss}_3[t_\mathrm{op},t_\mathrm{snk}]$ contributes (where we have assumed that
$t_\mathrm{src} = 0 < t_\mathrm{op} < t_\mathrm{snk} \leq N_t/2$). The spectral
decomposition for this three-point function is
\begin{equation}
  C^{cs\to ss}_3[t_\mathrm{op},t_\mathrm{snk}] = \sum_{i,j} B_{ji}
  e^{-E_j^{\eta_s} t_\mathrm{op} - E_i^{D_s} (t_\mathrm{snk}-t_\mathrm{op})}\,,
  \label{eq:specsemi}
\end{equation}
where we assumed around-the-world effects to be negligible and defined
\begin{equation}
  B_{ji} = \sqrt{A_j^{\eta_s}} M_{ji} \sqrt{A_i^{D_s}}\,,
\end{equation}
where $M_{ji} = \matrixel{{\eta_s}^j}{V_4}{{D_s}^i}$.  In order to compute the
nonperturbative form factor that parameterizes this decay, we need to extract the ground
state matrix element $M_{00}$ from fits to the two- and three-point functions. A common
approach is to construct ratios in which the temporal behavior cancels when the ordering
\mbox{$0 \ll t_\mathrm{op} \ll t_\mathrm{snk} \ll N_t/2$} is satisfied. One choice of such a
ratio is
\begin{equation}
  R[t_\mathrm{op},t_\mathrm{snk}] = \frac{C^{cs\to
  ss}_3[t_\mathrm{op},t_\mathrm{snk}]}{C_2^{ss}[t_\mathrm{op}]
  C_2^{cs}[t_\mathrm{snk}-t_\mathrm{op}]}\,,
  \label{eq:ratiosemi}
\end{equation}
which in the above limit approaches the value $M_{00}(A^{\eta_s}_0A^{D_s}_0)^{-\frac12}$.

In practice, the choice of $t_\mathrm{snk}$ is restricted by the temporal extent of the
lattice, so one has to carefully investigate whether a region for $t_\mathrm{op}$ and
$t_\mathrm{snk}$ can be found for which $R$ is approximately constant. The left-hand panel
of \cref{fig:semilep} shows this ratio as a function of $t_\mathrm{op}$ for 5 choices of
$t_\mathrm{snk}$. We find that even for the largest values of $t_\mathrm{snk}$, we cannot
isolate a region where stability in $t_\mathrm{op}$ and $t_\mathrm{snk}$ is observed,
indicating that for this dataset there is no region where a fit to a single state
is sufficient. Following our observations from the previous sections, a natural question
is whether applying suitably tuned Laplace filters to the data might allow for a
transformation of the data in which the excited states are sufficiently reduced to
identify a suitable time range where a single-state description of the data is
appropriate.

Since the spectrum of the correlation functions remains unchanged, separate Laplace
filters can be applied to any of the three correlation functions entering the construction
of \cref{eq:ratiosemi}. For the sake of simplicity, and because we find this to be
sufficient, we restrict ourselves to filtering the two-point functions that enter. The
right-hand panel of \cref{fig:semilep} shows the Laplace-filtered version of
\cref{eq:ratiosemi} where, prior to constructing the ratio, the two-point functions
$C_2^{cs}$ and $C_2^{ss}$ have been filtered using $\lambda^2_{D_s} = 2.9$ and
$\lambda^2_{\eta_s} = 1.2$, respectively. We stress that the asymptotic values these
ratios approach are different between the two panels. This is unsurprising since the
former should asymptotically plateau to $M_{00}(A^{\eta_s}_0A^{D_s}_0)^{-1/2}$, while for
the latter we have the replacement $A^{P}_0 \to {A_{\lambda_P;0}^{P}}$ for $P = \eta_s,
D_s$ (see \cref{eq:2ptfiltered}). However, from the same equation, the relationship
between $A_0$ and $A_{\lambda;0}$ is analytically known once $E_0$ has been determined, so
this does not pose any difficulty in practice. In \cref{fig:semilep_samescale}, we jointly
plot the original data (open symbols) together with the appropriately rescaled version of
the ratio. For this we use ground state energy estimates from late times of the effective
mass. Concretely, we use $E_0^{D_s} = 1.154$ and $E_0^{\eta_s} = 0.4030$. For a full
comparison between the filtered and the unfiltered result, the value for the energies
should be obtained from fits to the data and their uncertainties should be propagated.
However, this is not expected to make a large difference for the particular dataset at
hand since the statistical uncertainty on individual late-time effective mass data points
is below 2 per-mil in both cases.
From the plot we can draw several conclusions:
\begin{itemize}
  \item The filtered and unfiltered datasets approach the same asymptotic value, lending
    confidence that the method works.
  \item There are (few) data points for which the unfiltered dataset with
    $t_\mathrm{snk}=28$ is close
    to being ground-state dominated.
  \item For the filtered case, there are several values of $t_\mathrm{snk}$ which admit compatible plateaus for a
    range of time slices.
  \item The statistical uncertainties of the filtered data are of similar size to those of
    the unfiltered data.
\end{itemize}

We stress that for sufficiently precise data, the contamination from the first excited
state will always be larger than the statistical uncertainties, so the situation displayed
in the left panel of the figure \emph{necessarily} arises for large enough statistics. In
these situations, the Laplace filter provides a tunable parameter that can be adjusted to
eliminate the leading excited state (and the procedure can be iterated as long as the
contact term does not propagate into the time slices under consideration). Since the
spectral decomposition remains the same, the methodology carries over to other choices of
momenta and currents. The procedure can be further refined and extended by also filtering the
three-point function.

\subsection{Application to neutral meson mixing}

\begin{figure*}
  \includegraphics[width=\columnwidth]{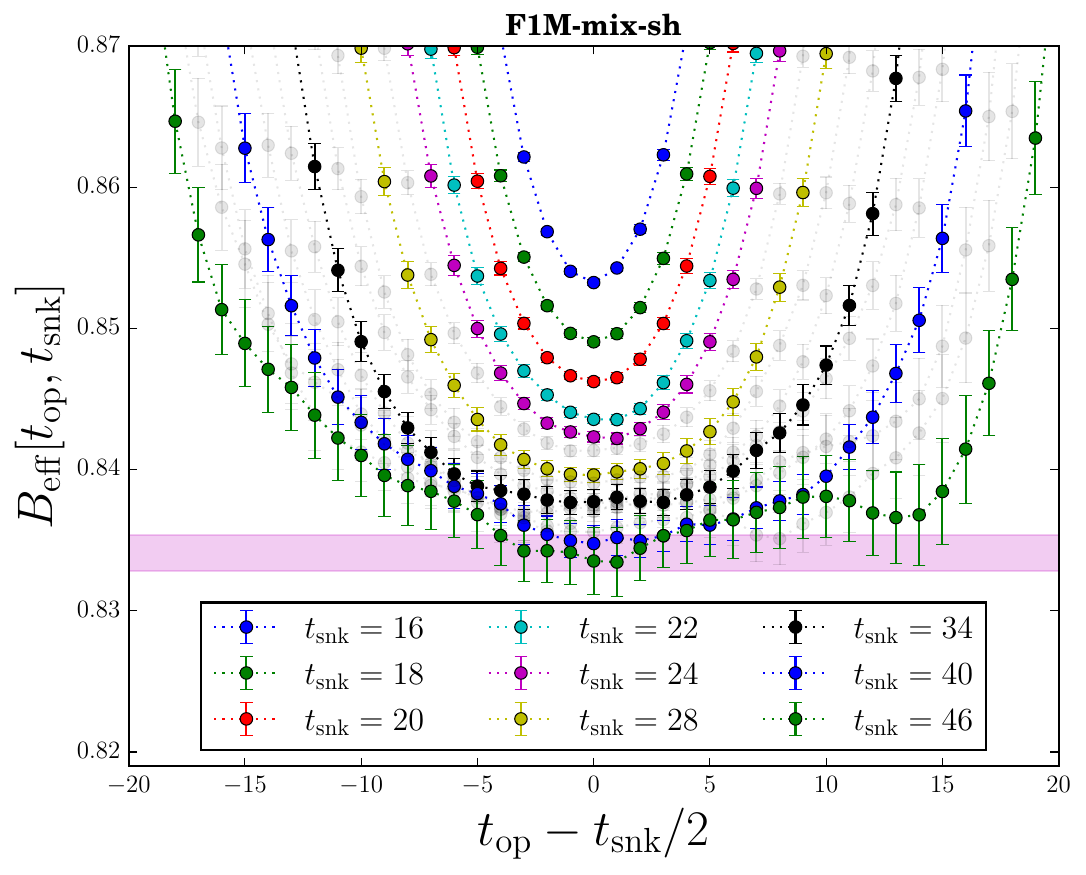}
  \includegraphics[width=\columnwidth]{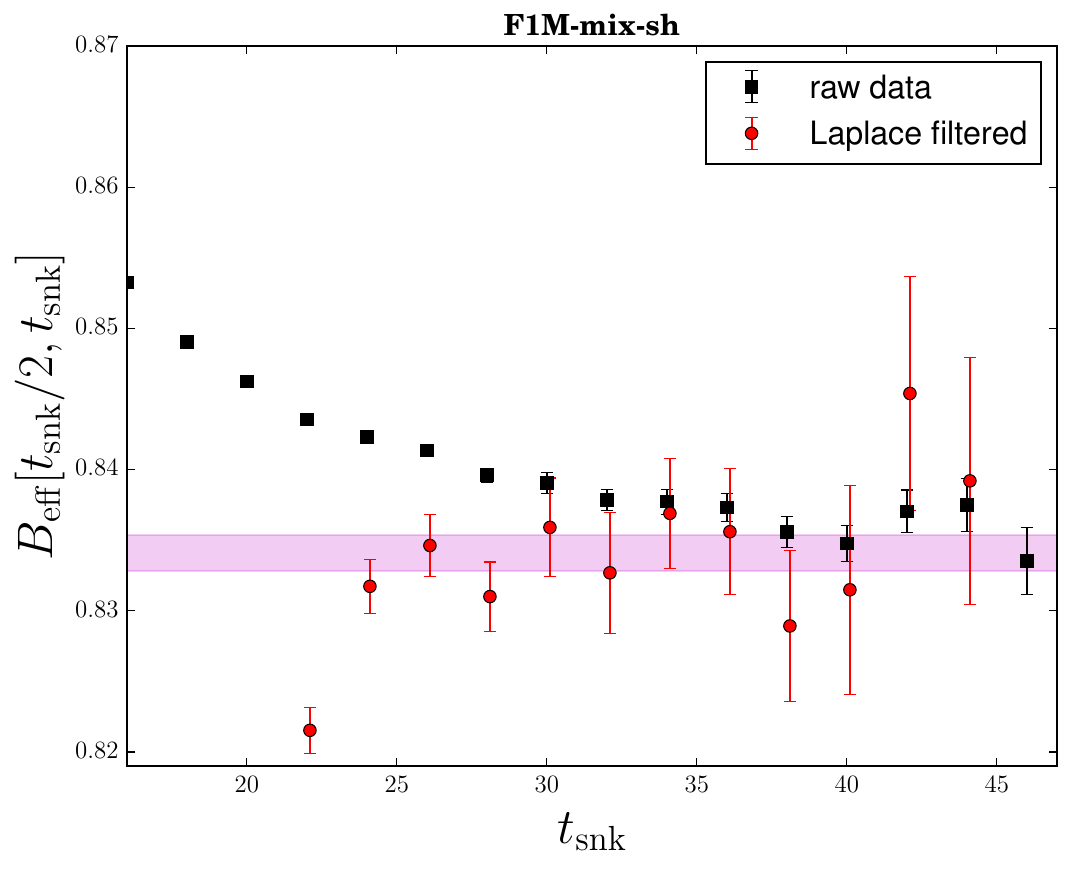}

  \caption{Analysis of the F1M-mix-sh dataset. The left-hand panel shows the effective
    bag parameter as a function of $t_\mathrm{op}$ for a variety of
    $t_\mathrm{snk}$ values.
    The right-hand plot shows the mid-point for the effective bag parameter between the
    source and the sink for the original data (black) and the Laplace filtered data (red)
    where the Laplace filter uses the values for the ground and first excited
    state energies
    obtained from a stability analysis analogous to that of
    Ref.~\citep{Boyle:2024gge}. The
    magenta band corresponds to the bag parameter as extracted from the same stability
  analysis.}
  \label{fig:mix}
\end{figure*}

The final dataset we consider describes neutral meson mixing. There are several
similarities between this and the case of semileptonic decays, but the required spectral
analysis is simplified by the fact that the initial and final state have the same
spectrum, so that \cref{eq:specsemi} becomes
\begin{equation}
  C^{sh}_3[t_\mathrm{op},t_\mathrm{snk}] = \sum_{i,j} D_{ij} e^{-E_j
  t_\mathrm{op} - E_i (t_\mathrm{snk}-t_\mathrm{op})}\,,
  \label{eq:specmix}
\end{equation}
with $D_{ji} = D_{ij}$. One immediate implication of this is that at fixed
$t_\mathrm{snk}$ the ground state contribution is a constant with respect to $t_\mathrm{op}$ rather than an exponential
decay. Analogous to the case of the semileptonic dataset (see \cref{subsec:semi}), we
define suitable ratios at fixed $t_\mathrm{snk}$ that allow for the comparison of
different values of $t_\mathrm{snk}$. However, since the temporal dependence with respect
to $t_\mathrm{op}$ cancels when the correlation function is ground-state dominated, the
three-point correlation functions are normalized by appropriate two-point functions at
single data points, such that the exponential suppression with $t_\mathrm{snk}$ is
removed. For historical reasons, the typical choice is to construct ratios
$B_\mathrm{eff}$ such that the asymptotic value approaches the so-called \emph{bag
parameter}, which is achieved by considering the two-point correlation functions $C^{PA}$
with $\gamma_\mathrm{src} = \gamma_5$ and $\gamma_\mathrm{snk} = \gamma_4
\gamma_5$.\footnote{More details can be found in
Refs.~\citep{Boyle:2018knm,Boyle:2024gge}.} The definition then reads
\begin{equation}
  B_\mathrm{eff}[t_\mathrm{op},t_\mathrm{snk}] =
  \frac{C_3[t_\mathrm{op},t_\mathrm{snk}]}{-C^{PA}_2[0]C^{PA}_2[t_\mathrm{snk}]}\,.
\end{equation}
Since the generation of this dataset does not require sequential sources and sources were
placed on every second time plane (see Refs.~\citep{Boyle:2018knm, Boyle:2024gge} for
details), many choices of $t_\mathrm{snk}$ are computationally accessible, and our dataset
comprises the values $t_\mathrm{snk} = 16,18,\cdots, 46$. The left-hand panel of
\cref{fig:mix} shows the effective bag parameter as a function of
$t_\mathrm{op}-t_\mathrm{snk}/2$ for different values of $t_\mathrm{snk}$. For better
readability, we do not display all values for $t_\mathrm{snk}$. Clearly, the data point
$t_\mathrm{op}=t_\mathrm{snk}/2$ (plotted as 0 on the x-axis) is furthest away from
both source and sink, and hence should have the smallest excited-state contamination. We
observe that for this data point, very small values of $t_\mathrm{snk}$ are statistically
incompatible with the larger values of $t_\mathrm{snk}$ and hence suffer from large
excited-state contamination. However, these data points are statistically the most precise, since the uncertainties grow as $t_\mathrm{snk}$ is increased.

An immediate question is whether Laplace filtering can be used to reduce the contamination
by excited states and hence leverage smaller values of $t_\mathrm{snk}$.
In the previous sections, we saw that the possibility to reduce excited states using our
choice of filter relies on the fact that the functional form of exponential decays is
unaltered. For that reason, rather than considering $C_3^{sh}$ at fixed $t_\mathrm{snk}$,
we consider the correlation function along the midpoint between the source at $t=0$ and
$t_\mathrm{snk}$. We then recover an exponential decay of the form
\begin{equation}
  C^{sh}_3[t_\mathrm{snk}/2,t_\mathrm{snk}] = \sum_{i,j} D_{ij} e^{-(E_i+E_j)
  t_\mathrm{snk}/2}\,,
  \label{eq:specmixmid}
\end{equation}
so that Laplace filtering is likely to be usable to remove unwanted excited states. A
comprehensive stability analysis analogous to that of Ref.~\citep{Boyle:2024gge} has been
performed to determine the ground and first excited state parameters of the two- and
three-point functions under consideration here, finding $E_0 = 0.94136(27)$, $E_1 =
1.190(21)$, and $B = 0.8341(13)$. Drawing on that knowledge, we apply a Laplace filter to
cancel the expected leading contamination. For the case of the mesonic two-point function
that enters the normalization, we choose $\lambda^2_\mathrm{mes} =
\tilde{E}_1^2$. For the three-point function, we note that the leading
excited-state contamination in \cref{eq:specmixmid} is $E_{01}=(E_0+E_1)/2$, and we therefore
apply a filter with $\lambda^2_\mathrm{mix} = \tilde{E}_{01}^2$. We then
construct the Laplace-filtered version of
$B_\mathrm{eff}[t_\mathrm{snk}/2,t_\mathrm{snk}]$. Given that we have knowledge of the
spectrum, we can also correct this value by the different normalization incurred from the
Laplace filtering and hence present the original and the filtered data on the same scale.
This is shown in the right-hand panel of \cref{fig:mix}. The black data points show the
original data as a function of $t_\mathrm{snk}$. There is clear evidence for excited-state
contamination for $t_\mathrm{snk} \lesssim 36$. For $t_\mathrm{snk} \gtrsim 36$ the
individual values start to be noisier, which can be understood by referring to the
left-hand plot. The red data points show the Laplace-filtered version of the data. As
expected, we observe that already for $t_\mathrm{snk}\approx 26$ we obtain ground-state
dominated data with similar precision to those of the earliest ground-state dominated data
points of the unfiltered data. We further find that different choices of $t_\mathrm{snk}$
fluctuate more for the filtered data, but this is at least partially by virtue of the
achieved decorrelation from using a Laplace filter and hence the data points carry more
individual information.

While this particular dataset at the current level of statistical uncertainties allows for
parameter choices with ground-state dominance, there are many examples where this would
not be the case, for example if the statistics were significantly increased, or for
operators with larger excited-state contamination and/or ensembles with smaller temporal
extent.

\section{Discussion and conclusion}
In this paper, we have addressed two major issues that lattice analysis workflows
frequently encounter: the potential ill-conditioning of data-estimated correlation
matrices and the extraction of the spectrum from correlation functions in the presence of
excited state contamination. Applying a transformation to the data and the model does not
involve any ad hoc modifications of the spectrum of the correlation matrix and can be used
to ameliorate the first problem. As one such transformation, we advocate the use of regulated Laplace
filters, as they are linear and invertible, and preserve the exponential form typically
encountered in correlation functions.

We find that a suitably chosen Laplace filter significantly improves the condition number
of the correlation matrix, and that down-sampling positively interacts with this, \ie, the
two effects are additive. This holds in all investigated examples, and a wide parameter
range in the regulator $\lambda$ allows for a reduction of the condition number.

A key observation is that the application of a Laplace filter modifies the amplitudes of
all states $E_i$ by a factor $(\lambda^2-\tilde{E}_i^2)$. This can be exploited in
multiple ways:

\begin{enumerate}
  \item Finding the values of $\lambda_i$ that eliminate the correlation
    function in a given
    time range can be used to obtain the spectrum without any need to perform non-linear
    multi-exponential fits and without needing to fit the corresponding
    amplitudes, halving
    the number of parameters that need to be fitted. We find this method to be robust and
    stable.
  \item Extending the above, we devised a method to determine the number of statistically
    significant states in a given time range. This can be used to choose the appropriate
    ansatz for a traditional multi-state fit. Combined with the previous point,
    this method
    can provide data-driven priors or initial guesses for fit parameters.
  \item By tuning $\lambda$, we can eliminate unwanted excited state contamination. We
    demonstrated this application for typical state-of-the-art correlation
    functions. Applying
    this method allows fits to earlier and hence statistically more precise
    time slices. We
    particularly envisage applications to correlation functions with poor signal-to-noise
    properties and correlation functions that require multiple well-separated
    timescales. A
    further advantage is that traditional analysis frameworks can be used to
    perform the fits,
    making the method easy to implement and use.
\end{enumerate}
We reiterate that Laplace filtering can be applied in conjunction with the various methods discussed in the Introduction should the resulting correlation matrix not yet be sufficiently well conditioned.

We envisage several further applications of this work in the future. We will apply this
method to various ongoing large-scale numerical efforts, such as semileptonic decays and
neutral meson mixing. Furthermore, it can be employed to investigate and scrutinize
excited state effects, such as those frequently observed in nucleon matrix
elements~\citep{Bar:2021crj, Bar:2018xyi} and those expected in the heavy-light mesonic
sector~\citep{Bar:2023sef}. This method could also be of great value to finite temperature simulations where the temporal extent has to be small by construction and hence excited state contamination is a big concern. In particular for anisotropic simulations with many points in the temporal direction several Laplace filters could be applied without propagating the contact term into the region of interest.

Finally, we plan to extend the use of the method as a novel way of extracting the spectrum
without the need for multi-exponential fits by rigorously attaching statistical
uncertainties to the determination of energies and amplitudes obtained from such a maximum
likelihood determination.

\begin{acknowledgments}
The authors would like to gratefully thank to our collaborators within the RBC/UKQCD
collaboration, in particular Felix Erben and Andreas J\"uttner, for the use of the
ensembles and the data presented in this study. AP received funding from the European
Research Council (ERC) under the European Union's Horizon 2020 research and innovation
programme under grant agreements No 757646 \& 813942. AP is additionally funded in part by
UK STFC grant ST/X000494/1. AP was additionally supported by a CERN Scientific Associate
position, as well as by a Long Term Invitational Fellowship from the Japan Society for the
Promotion of Science. This work used the DiRAC Extreme Scaling service (Tursa / Tesseract) at the University of Edinburgh, managed by the EPCC
on behalf of the STFC DiRAC HPC Facility (\url{ www.dirac.ac.uk}). The DiRAC service at
Edinburgh was funded by BEIS, UKRI and STFC capital funding and STFC operations grants.
DiRAC is part of the UKRI Digital Research Infrastructure. Two-point correlators
used in this work are publicly available at \citep{tsang_2025_16921526}, three-point functions
are still part of ongoing projects and are available upon reasonable request from the authors.
\end{acknowledgments}
\bibliography{laplace-filter}
\end{document}